\documentclass[12pt]{article}
\usepackage{amsthm, bm, graphicx, hyperref, mathrsfs, bbm}
\usepackage{amsfonts}
\usepackage{amsmath}
\usepackage{amssymb}
\usepackage{dsfont}
\usepackage{graphicx}
\usepackage{color}
\usepackage{braket}
\usepackage{epstopdf}
\usepackage[footnotesize]{caption}
\usepackage{bigfoot}
\usepackage{amsthm}
\usepackage{enumitem}
\usepackage{mathrsfs}
\usepackage{mathtools}     
\usepackage{subeqnarray}         
\usepackage{cases}               
\usepackage{color}
\usepackage{subfigure}
\usepackage{cite}               
\usepackage{hyperref}            
\usepackage{multirow,makecell}   
\usepackage{textcomp}
\usepackage{wasysym}
\usepackage{url}
\usepackage[margin=3cm]{geometry}
\usepackage{geometry}
\usepackage{amsfonts}
\usepackage{graphicx}
\usepackage{float}
\usepackage{amsmath}
\usepackage{hyperref}
\usepackage{tensor}
\usepackage{subfigure}
\usepackage{cases}
\usepackage{appendix}
\usepackage{multirow}
\usepackage{color}
\allowdisplaybreaks[3]

\newcommand*{\dif}{\mathop{}\!\mathrm{d}}
\DeclareNewFootnote{default} 
\DeclareNewFootnote{B}

\begin{document}
\begin{titlepage}
\vspace{0.5cm}
\begin{center}
{\Large\bf{Holographic Correlators of Boundary/Crosscap CFTs in Two Dimensions}}
\lineskip .75em
\vskip 1.5cm
{\large{Yun-Ze Li\footnoteB{These authors contributed equally to the work.\label{coauthor}}$^{,a,}$\footnote{lyz21@mails.jlu.edu.cn}, Yunfei Xie\textsuperscript{,\ref{coauthor}}$^{,a,}$\footnote{jieyf22@mails.jlu.edu.cn}, Song He$^{b,c,a,d,}$\footnote{hesong@nbu.edu.cn (Corresponding author)}}}
\vskip 2.5em
{\normalsize\it $^a$Center for Theoretical Physics and College of Physics, Jilin University,\\
 Changchun 130012, People's Republic of China
\\
$^{b}$Institute of Fundamental Physics and Quantum Technology, \\ Ningbo University, Ningbo, Zhejiang 315211, China \\
$^{c}$School of Physical Science and Technology, 
Ningbo University,\\ Ningbo, 315211, China 
\\$^d$Max Planck Institute for Gravitational Physics (Albert Einstein Institute),\\ 
Am M\"uhlenberg 1, 14476 Golm, Germany}
\end{center}
\begin{abstract}
This work explores holographic correlators within the frameworks of two-dimensional Boundary Conformal Field Theory (BCFT) and Crosscap Conformal Field Theory (XCFT). Utilizing the AdS/CFT correspondence, we compute stress tensor correlators in BCFT, considering both tensionless and tensionful end-of-the-world (EOW) brane scenarios. We derive recurrence relations for two-point and three-point correlators and examine the impact of non-zero brane tension on correlators. Extending these results, we investigate the holographic duals of XCFTs, presenting explicit scalar and stress tensor correlator computations on projective geometries such as $\mathbb{RP}^2$. Additionally, we analyze stress tensor correlators at a finite cutoff, uncovering deformations to one-point and two-point functions induced by the cutoff. Our findings provide novel insights into the holographic structures of BCFT and XCFT while laying the groundwork for future research into higher-dimensional extensions.
\end{abstract}
\end{titlepage}
\newpage
\tableofcontents
\section{Introduction}
The Anti-de Sitter gravity/conformal field theory (AdS/CFT) correspondence \cite{Maldacena:1997re, Gubser:1998bc, Witten:1998qj, Aharony:1999ti}, as a concrete realization of the holographic principle \cite{tHooft:1993dmi, Susskind:1994vu}, provides a powerful tool for analytically studying strongly coupled quantum field theories. One significant application of this correspondence lies in obtaining the correlators of local operators in the dual CFT by performing gravitational perturbative calculations in bulk. Stress tensor correlators contain critical information about a system's energy, momentum, and stress distribution, facilitating analysis of phenomena such as the c-theorem \cite{Zamolodchikov:1986gt}. The holographic calculations of stress tensor correlators have been studied in many remarkable works \cite{Liu:1998ty, Arutyunov:1999nw, DHoker:1999bve, Bianchi:2001de, Bianchi:2001kw, Papadimitriou:2004rz, Raju:2012zs, Bagchi:2015wna, Nguyen:2021pdz}. In our previous works, we have computed the holographic stress tensor correlators on torus \cite{He:2023hoj, He:2023knl, He:2024fdm}, and higher genus Riemann surfaces \cite{He:2024xbi} within the framework of AdS$_3$/CFT$_2$. These calculations have been extended to AdS$_5$/CFT$_4$ in \cite{He:2023wcs}. This paper aims to investigate the holographic stress tensor correlators of two-dimensional boundary conformal field theories (BCFT$_2$) and CFTs on the real projective plane, as these two cases exhibit analogous bulk constructions.\par
The AdS/BCFT correspondence \cite{Karch:2000gx, Takayanagi:2011zk, Fujita:2011fp, Nozaki:2012qd} is a generalized version of  AdS/CFT correspondence. The basic idea is to extend the boundary of BCFT into the bulk to form an end-of-the-world (EOW) brane while imposing the Neumann boundary condition on it\footnote{The EOW brane with a Dirichlet or mixed boundary condition has been studied in \cite{Miao:2018qkc, Guijosa:2022jdo, Miao:2017gyt, Chu:2017aab}.}. One advantage of AdS/BCFT correspondence is that the calculation of holographic entanglement entropy \cite{Ryu:2006bv, Ryu:2006ef, Hubeny:2007xt} in this setup is straightforward. Furthermore, holographic correlators in AdS/BCFT can be computed by the standard Gubser-Klebanov-Polyakov-Witten (GKPW) relation \cite{Gubser:1998bc, Witten:1998qj}. Recently, holographic correlators of primary operators have been extensively investigated using various methods \cite{Alishahiha:2011rg, Setare:2012ks, Hinterbichler:2015pta, Kastikainen:2021ybu, Park:2024pkt, Tian:2024fmo}. This study focuses on holographic correlators of the stress tensor in AdS$_3$/BCFT$_2$. One notable distinction is that the brane bending \cite{Izumi:2022opi, Suzuki:2022yru} should be considered when calculating stress tensor correlators. We first examine the correlators in the Poincare AdS$_3$ background with a tensionless brane. The Neumann boundary condition on the brane determines the brane profile. From the field theory perspective, it constrains the boundary values of stress tensor correlators, corresponding to the Cardy condition \cite{Cardy:1984bb, Cardy:2004hm}. More interestingly, when we switch to the hyperbolic slicing coordinates, the Neumann boundary condition no longer provides boundary conditions for stress tensor correlators; instead, it relates these correlators to the variations of the brane profile. For instance, in the calculation of two-point correlators, we find
\begin{align}
	\frac{\delta T_{ij}(t,x)}{\delta g_{kl}(t_0,x_0)}=\widehat{L}_{ij}\Big(\frac{\delta \psi(t,x)}{\delta g_{kl}(t_0,x_0)}\Big)+\text{contact terms}.\label{2-pt in terms of brane profile}
\end{align}
Here, $\psi(t,x)$ is the brane profile, and $\widehat{L}_{ij}$ is a second-order differential operator. In this expression, the conservation law of the stress tensor is automatically fulfilled, while the Weyl anomaly yields a differential equation governing the brane profile.\par
In addition to BCFTs, conformal field theories on non-orientable manifolds have also received significant attention due to research on non-orientable strings \cite{Burgess:1986ah, Ishibashi:1988kg, Fioravanti:1993hf, Pradisi:1995pp, Giombi:2020xah, Tsiares:2020ewp}. From the perspective of field theory, the crosscap correlators of the compactified boson CFT and the Ising CFT have been investigated in \cite{Tan:2024dcd} and \cite{Zhang:2024rnh}, respectively. Lots of studies have been dedicated to establishing the holographic dual of non-orientable CFTs \cite{Verlinde:2015qfa, Nakayama:2016xvw, Lewkowycz:2016ukf, Maloney:2016gsg, Caetano:2022mus, Wei:2024zez, Wei:2024kkp}. In this paper, we compute holographic scalar correlators and stress tensor correlators on the real projective plane $\mathbb{RP}^2$ based on the bulk construction presented in \cite{Wei:2024zez}. This construction introduces EOW branes in bulk to resolve the singularities that arise from the $\mathbb{Z}_2$ quotient operation, calculating correlators similar to the hyperbolic slicing case in AdS$_3$/BCFT$_2$.\par
This paper is organized as follows: In Section 2, we present the holographic setup for calculating stress tensor correlators in Boundary Conformal Field Theory, including both the tensionless and tensionful cases, and discuss the associated recurrence relations. Section 3 extends these results to the holographic dual of the crosscap Conformal Field Theory, where we analyze scalar and stress tensor correlators, focusing on configurations involving projective spaces. Finally, in Section 4, we summarize the main findings of this work and outline possible avenues for future research.

\section{Holographic stress tensor correlators of BCFT$_2$}\label{section 2}
In this section, we compute the holographic correlators of stress tensor within the framework of AdS$_3$/BCFT$_2$ correspondence. Our calculations are performed in the semiclassical limit. We first consider the case of tensionless brane and calculate the two-point and three-point stress tensor correlators in the Poincare AdS coordinates. We also derive an explicit recurrence relation of the higher-point correlators. Subsequently, we employ hyperbolic slicing coordinates to investigate the stress tensor correlators when the brane has a non-zero tension.
\subsection{Holographic setup and stress tensor correlators}
Let us briefly review the fundamentals of AdS$_3$/BCFT$_2$ correspondence. Suppose the field theory we are concerned with lives on a two-dimensional manifold $M$ with a boundary $\partial M$. When considering the gravity dual of the field theory, the boundary $\partial M$ extends into the bulk and forms a two-dimensional brane $Q$. Then, the boundary of the three-dimensional bulk $\mathcal{M}$ consists of two parts, 
\begin{align}
    \partial\mathcal{M}=M\cup Q\ \ \text{with}\ \ \partial Q=\partial M.
\end{align}\par
In this section, we focus on calculating the holographic stress tensor correlators. We consider a pure gravitational system in the bulk, where the boundary metric is the only source in the corresponding CFT. The total bulk action consists of three parts,
\begin{align}
    I_{\text{bulk}}=-\frac{1}{16\pi G}\int_{\mathcal{M}}\text{d}^3x\sqrt{\mathcal{G}}\Big(\mathcal{R}+2\Big)+I_{M}+I_{Q}.\label{AdS3/BCFT2 bulk action}
\end{align}
The first term is the Einstein-Hilbert action with a negative cosmological constant\footnote{We have set the AdS radius $l=1$.}. The boundary term $I_{M}$ contains both the Gibbons-Hawking term \cite{Gibbons:1976ue} and the counter term \cite{deHaro:2000vlm} (see also \cite{Henningson:1998gx,Bianchi:2001kw,Skenderis:2002wp,Balasubramanian:1999re,Emparan:1999pm,Kraus:1999di}),
\begin{align}
    I_{M}=-\frac{1}{8\pi G}\int_{M}\text{d}^2x\sqrt{\gamma}\Big(K-1\Big),
\end{align}
where $K=\gamma^{ij}K_{ij}$ with the induced metric $\gamma_{ij}$ and the extrinsic curvature $K_{ij}$. The boundary term on the EOW brane $Q$ takes the form
\begin{align}
    I_{Q}=-\frac{1}{8\pi G}\int_{Q}\text{d}^2x\sqrt{\gamma}\Big(K-T\Big).
\end{align}
The constant $T$ represents the tension of the EOW brane. In the gravitational calculation, two boundaries $M$ and $Q$ have different prescriptions. For the conformal boundary $M$, we impose the Dirichlet boundary condition $\delta\gamma_{ij}|_{M}=0$, and the boundary metric serves as the source of the stress tensor operator in CFT$_2$. In contrast, the Neumann boundary condition is imposed on the EOW brane \cite{Takayanagi:2011zk, Fujita:2011fp},
\begin{align}
    (K_{ij}-K\gamma_{ij}+T\gamma_{ij})\Big|_{Q}=0.\label{Neumann boundary condition BCFT2}
\end{align}
The EOW brane $Q$ becomes dynamical with the constraint of (\ref{Neumann boundary condition BCFT2}), a feature that will play a crucial role in the subsequent computations of stress tensor correlators.\par
It is convenient to work in the Fefferman-Graham coordinates \cite{Fefferman:1985ci,Fefferman:2007rka}, in which the bulk metric $\mathcal{G}_{\mu\nu}$ takes the form 
\begin{align}
    \mathcal{G}_{\mu\nu}\text{d}x^{\mu}\text{d}x^{\nu}=\frac{\text{d}z^2}{z^2}+\frac{1}{z^2}g_{ij}(z,\boldsymbol{x})\text{d}x^{i}\text{d}x^{j}.\label{Fefferman Graham coordinates}
\end{align}
Here, $z$ is the radial coordinate, and the conformal boundary is at $z=0$. The bulk solution $g_{ij}(z,\boldsymbol{x})$ can be constructed near the conformal boundary. For a pure gravitational system in 3D space specific saddle point dominates the partition functions as \cite{Banados:1998gg,Skenderis:1999nb}
\begin{align}
    g_{ij}(z,\boldsymbol{x})=g_{(0)ij}(\boldsymbol{x})+z^2g_{(2)ij}(\boldsymbol{x})+z^4g_{(4)ij}(\boldsymbol{x}).\label{Fefferman-Graham expansion}
\end{align}
In the Fefferman-Graham coordinates, the bulk Einstein's equation is reduced to the following three equations:
\begin{align}
    g_{(4)ij}&=\frac{1}{4}g_{(2)ik}g_{(0)}^{kl}g_{(2)lj},\label{g4 expression}\\
    \nabla_{(0)}^{i}g_{(2)ij}&=\nabla_{(0)j}(g^{kl}_{(0)}g_{(2)kl}),\label{conservation equation 0}\\
    g^{kl}_{(0)}g_{(2)kl}&=-\frac{1}{2}R_{(0)}\label{trace relation 0},
\end{align}
where $\nabla_{(0)}$ and $R_{(0)}$ indicate the covariant derivative operator and the Ricci scalar of $g_{(0)}$ respectively.\par
Throughout this paper, we employ the standard GKPW relation \cite{Gubser:1998bc, Witten:1998qj} to compute holographic correlators, which establishes the equivalence between the bulk gravitational partition function and the boundary-generating functional,
\begin{align}
    Z_{\text{G}}[\phi_{(0)},g_{(0)ij}]=\Big\langle{\text{exp}\Big[\int_{\partial\mathcal{M}}d^2x\sqrt{g_{(0)}}\Big(\phi_{(0)}O-\frac{1}{2}g^{ij}_{(0)}T_{ij}}\Big)\Big]\Big\rangle_{\text{CFT}}.\label{GKPW rekation}
\end{align}
In the semiclassical limit, the gravitational partition function can be approximated as a sum over all classical saddle points, $Z_{\text{G}}\approx\sum_{\alpha}e^{-I^{(\alpha)}_{\text{on-shell}}}$. Assuming that the partition function is dominated by a specific saddle point, holographic correlators can be obtained from the functional derivatives of its on-shell action. In particular, the stress tensor correlators can be written as
\begin{align}
    &\Big\langle{\prod_{n=1}^N T_{i_nj_n}(\boldsymbol{x}_n)}\Big\rangle_{c}=-\Big(\prod_{n=1}^N\frac{-2}{\sqrt{g_{(0)}(\boldsymbol{x}_n)}}\Big)\Big(\prod_{n=1}^N\frac{\delta}{\delta g^{i_nj_n}_{(0)}(\boldsymbol{x}_n)}\Big)I_{\text{on-shell}}\Big|_{g_{(0)ij}=\eta_{ij}} ,\label{holographic correlator}
\end{align}
where the subscript $c$ implies the connected part of the correlator. Among them, the one-point correlator corresponds to the Brown-York tensor \cite{Brown:1992br},
\begin{align}
    \langle{T_{ij}}\rangle&=-\frac{1}{8\pi G}(K_{ij}-K\gamma_{ij}+\gamma_{ij})\Big|_{M}\notag\\
    &=\frac{1}{8\pi G}(g_{(2)ij}+\frac{1}{2}R_{(0)}g_{(0)ij}). \label{stress tensor one-point correlators}
\end{align}
The Newton's constant $G$ is related to the CFT central charge through the Brown-Henneaux relation \cite{Brown:1986nw} $c=\frac{3}{2G}$. Plugging (\ref{stress tensor one-point correlators}) into (\ref{conservation equation 0}) and (\ref{trace relation 0}), we obtain the conservation law and the trace relation,
\begin{align}
\nabla_{(0)}^j\langle{T_{ij}}\rangle&=0,\label{conservation equation 1}\\
    g^{ij}_{(0)}\langle{T_{ij}}\rangle&=\frac{1}{16\pi G}R_{(0)}.\label{trace relation 1}
\end{align}\par
For most of this section, our calculations are performed in the Poincare AdS$_3$ background,
\begin{align}
    \dif s^2=\frac{\dif z^2+\dif t^2+\dif x^2}{z^2}.\label{Poincare AdS3 metric}
\end{align}
The field theory is defined on the right half plane $x\geq 0$. The Poincare AdS$_3$ is dual to the vacuum state in a CFT$_2$. Other gravitational saddle points, which are dual to excited states, can be obtained from the Poincare AdS$_3$ by coordinate transformations \cite{Banados:1998gg, Roberts:2012aq}. Constrained by the Neumann boundary condition (\ref{Neumann boundary condition BCFT2}), the profile of the EOW brane in the background (\ref{Poincare AdS3 metric}) is
\begin{align}
    Q:\ \ \ \ x=kz,
\end{align}
where the constant $k$ is determined by the brane tension through $T=-\frac{k}{\sqrt{1+k^2}}$.
\subsection{Tensionless case}
We start by placing a tensionless brane at $x=0$ in the Poincare AdS$_3$. To compute the stress tensor correlators, one must perturb the boundary metric and solve the bulk Einstein's equation order by order.
\subsubsection{Two-point and three-point correlators}
 Consider the following variation
\begin{align}
    g_{(0)ij}(\epsilon;\boldsymbol{x})=\eta_{ij}+\epsilon \chi_{ij}(\boldsymbol{x}),\label{metric perturbation}
\end{align}
where $\epsilon$ is an infinitesimal parameter. The higher-order Fefferman-Graham coefficients can be written as series expansions in $\epsilon$,
\begin{align}
    g_{(2)ij}(\epsilon;\boldsymbol{x})&=\sum_{n=1}^{\infty}\epsilon^{n}g^{[n]}_{(2)ij}(\boldsymbol{x}),\notag\\
    g_{(4)ij}(\epsilon;\boldsymbol{x})&=\frac{1}{4}[g_{(2)ik}g_{(0)}^{kl}g_{(2)lj}](\epsilon;\boldsymbol{x})=\sum_{n=1}^{\infty}\epsilon^{n}g^{[n]}_{(4)ij}(\boldsymbol{x}).\label{bulk metric perturbation}
\end{align}
The coefficient $g_{(2)ij}$ can be expressed in terms of the boundary metric and the Brown-York tensor using definition (\ref{stress tensor one-point correlators}). To first order in $\epsilon$, we have
\begin{align}
    g^{[1]}_{(2)tt}&=8\pi G\langle{T_{tt}}\rangle^{[1]}+\frac{1}{2}(\partial_{x}^2\chi_{tt}-2\partial_t\partial_x\chi_{tx}+\partial_{t}^2\chi_{xx}),\notag\\
   g^{[1]}_{(2)tx}&=8\pi G\langle{T_{tx}}\rangle^{[1]},\notag\\
 g^{[1]}_{(2)xx}&=8\pi G\langle{T_{xx}}\rangle^{[1]}+\frac{1}{2}(\partial_{x}^2\chi_{tt}-2\partial_t\partial_x\chi_{tx}+\partial_{t}^2\chi_{xx}).\label{g21 definition}
\end{align}
From (\ref{conservation equation 1}) and (\ref{trace relation 1}) $\langle{T_{ij}}\rangle^{[1]}$ is constrained by 
\begin{align}
\partial_t\langle{T_{tt}}\rangle^{[1]}+\partial_{x}\langle{T_{tx}}\rangle^{[1]}&=\partial_{x}\langle{T_{xx}}\rangle^{[1]}+\partial_{t}\langle{T_{tx}}\rangle^{[1]}=0,\label{2-pt Tensionless Conservation equation}\\
    \langle{T_{tt}}\rangle^{[1]}+\langle{T_{xx}}\rangle^{[1]}&=-\frac{1}{16\pi G}(\partial_{x}^2\chi_{tt}-2\partial_t\partial_x\chi_{tx}+\partial_{t}^2\chi_{xx}).\label{2-pt Tensionless Trace relation}
\end{align}
One can easily find that $\langle{T_{tx}}\rangle^{[1]}$ satisfies the Laplacian equation,
\begin{align}
(\partial_t^2+\partial_x^2)\langle{T_{tx}}\rangle^{[1]}=\frac{1}{16\pi G}(\partial_t\partial_x^3\chi_{tt}-2\partial_t^2\partial_x^2\chi_{tx}+\partial_t^3\partial_x\chi_{xx}). \label{Laplacian equation}
\end{align}
Meanwhile, the perturbed bulk geometry should fulfill the Neumann boundary condition on the EOW brane. As we perturb the boundary metric, it induces variations in the profile of the EOW brane, which can be formally expressed as
\begin{align}
    Q:\ \ \ x(\epsilon;z,t)=\sum_{n=0}^{\infty}\epsilon^nf^{[n]}(z,t). \label{tensionless brane profile}
\end{align}
For the tensionless case we have $f^{[0]}(z,t)=0$. Plugging (\ref{g21 definition}) and (\ref{tensionless brane profile}) into the Neumann boundary condition (\ref{Neumann boundary condition BCFT2}), we obtain
\begin{align}
    &\frac{(z\partial_{z}^2-\partial_z)f^{[1]}(z,t)}{z^2}=0,\notag\\
   &8\pi G[\langle{T_{tx}}\rangle^{[1]}](t,0)+\frac{\partial_z\partial_tf^{[1]}(z,t)}{z}=0,\notag\\
    &8\pi Gz[\partial_t\langle{T_{tx}}\rangle^{[1]}](t,0)-4\pi Gz[\partial_x\langle{T_{tt}}\rangle^{[1]}](t,0)+\frac{(z\partial_{t}^2-\partial_z)f^{[1]}(z,t)}{z^2}\notag\\
    &\quad\quad-\frac{[\partial_x\chi_{tt}-2\partial_t\chi_{tx}](t,0)}{2z}-\frac{z}{4}[\partial_x^3\chi_{tt}-2\partial_t\partial_x^2\chi_{tx}+\partial_t^2\partial_x\chi_{xx}](t,0)=0.\label{2-pt Tensionless NBC}
\end{align}
By combining (\ref{2-pt Tensionless Conservation equation})(\ref{2-pt Tensionless Trace relation})(\ref{2-pt Tensionless NBC}), we find
\begin{align}
    f^{[1]}(z,t)&=-\frac{z^2}{4}[\partial_x\chi_{tt}-2\partial_t\chi_{tx}](t,0),\label{2-pt tensionless brane profile}\\
    \langle{T_{tx}(t,0)}\rangle^{[1]}&=\frac{1}{16\pi G}[\partial_t\partial_x\chi_{tt}-2\partial_t^2\chi_{tx}](t,0).\label{Tensionless boundary condition of Ttx}
\end{align}
In addition to (\ref{Tensionless boundary condition of Ttx}), we still require an additional boundary condition to obtain the exact solution of equation (\ref{Laplacian equation}). This boundary condition arises from the regularity of the two-point correlators $\frac{\delta\langle{T_{tx}(\boldsymbol{x})}\rangle^{[1]}}{\delta\chi_{ij}(\boldsymbol{x}_0)}$. For a finite $\boldsymbol{x}_0$, we allow $x$ to approach infinity while ensuring that the value of two-point correlators remains finite,
\begin{align}
    \frac{\delta\langle{T_{tx}(\boldsymbol{x})}\rangle^{[1]}}{\delta\chi_{ij}(\boldsymbol{x}_0)}\Big|_{x\to\infty}\ \ \ \ \text{finite}.
\end{align}
Putting everything together, we obtain the two-point correlators $\langle{T_{tx}T_{ij}}\rangle$,
\begin{align}
    \langle{T_{tx}(\boldsymbol{x})T_{tt}(\boldsymbol{x}_0)}\rangle&=\frac{3i}{16\pi^2 G}\Big[\frac{1}{(t-t_0+i(x+x_0))^4}-\frac{1}{(t-t_0-i(x+x_0))^4}\notag\\
    &\quad+\frac{1}{(t-t_0+i(x-x_0))^4}-\frac{1}{(t-t_0-i(x-x_0))^4}\Big]\notag\\
    &\quad+\frac{1}{8\pi G}\partial_{t}\partial_{x}\delta(\boldsymbol{x}-\boldsymbol{x_0}),\notag\\
    \langle{T_{tx}(\boldsymbol{x})T_{tx}(\boldsymbol{x}_0)}\rangle&=\frac{3}{16\pi^2 G}\Big[\frac{1}{(t-t_0+i(x+x_0))^4}+\frac{1}{(t-t_0-i(x+x_0))^4}\notag\\
    &\quad-\frac{1}{(t-t_0+i(x-x_0))^4}-\frac{1}{(t-t_0-i(x-x_0))^4}\Big]\notag\\
    &\quad-\frac{1}{8\pi G}\partial_{t}^2\delta(\boldsymbol{x}-\boldsymbol{x_0}),\notag\\
     \langle{T_{tx}(\boldsymbol{x})T_{xx}(\boldsymbol{x}_0)}\rangle&=\frac{-3i}{16\pi^2 G}\Big[\frac{1}{(t-t_0+i(x+x_0))^4}-\frac{1}{(t-t_0-i(x+x_0))^4}\notag\\
    &\quad+\frac{1}{(t-t_0+i(x-x_0))^4}-\frac{1}{(t-t_0-i(x-x_0))^4}\Big],\label{tensionless 2-pt result 1}
\end{align}
where $\delta(\boldsymbol{x}-\boldsymbol{x}_0)=\delta(t-t_0)\delta(x-x_0)$. The other two-point correlators can be determined by equations (\ref{2-pt Tensionless Conservation equation}) and (\ref{2-pt Tensionless Trace relation}), along with the Bose symmetry of the correlators,
\begin{align}
    \langle{T_{tt}(\boldsymbol{x})T_{tt}(\boldsymbol{x}_0)}\rangle&=\frac{3}{16\pi^2 G}\Big[\frac{1}{(t-t_0+i(x+x_0))^4}+\frac{1}{(t-t_0-i(x+x_0))^4}\notag\\
    &\quad+\frac{1}{(t-t_0+i(x-x_0))^4}+\frac{1}{(t-t_0-i(x-x_0))^4}\Big]\notag\\
    &\quad+\frac{1}{8\pi G}(\partial_{t}^2-\partial_{x}^2)\delta(\boldsymbol{x}-\boldsymbol{x}_0),\notag\\
    \langle{T_{tt}(\boldsymbol{x})T_{xx}(\boldsymbol{x}_0)}\rangle&=-\frac{3}{16\pi^2 G}\Big[\frac{1}{(t-t_0+i(x+x_0))^4}+\frac{1}{(t-t_0-i(x+x_0))^4}\notag\\
    &\quad+\frac{1}{(t-t_0+i(x-x_0))^4}+\frac{1}{(t-t_0-i(x-x_0))^4}\Big]\notag\\
    &\quad-\frac{1}{8\pi G}\partial_{t}^2\delta(\boldsymbol{x}-\boldsymbol{x}_0),\notag\\
    \langle{T_{xx}(\boldsymbol{x})T_{xx}(\boldsymbol{x}_0)}\rangle&=\frac{3}{16\pi^2 G}\Big[\frac{1}{(t-t_0+i(x+x_0))^4}+\frac{1}{(t-t_0-i(x+x_0))^4}\notag\\
    &\quad+\frac{1}{(t-t_0+i(x-x_0))^4}+\frac{1}{(t-t_0-i(x-x_0))^4}\Big].\label{tensionless 2-pt result 2}
\end{align}
As a special case, we have
\begin{align}
	\langle{T_{xx}(t,0)T_{xx}(t_0,0)}\rangle=\frac{3}{4\pi^2G(t-t_0)^4}=\frac{c}{2\pi^2(t-t_0)^4},
\end{align}
which is consistent with the result in \cite{McAvity:1993ue}.\par
At the second order in $\epsilon$, the trace relation and the conservation law take the forms
\begin{align}
    \langle{T_{tt}}\rangle^{[2]}+\langle{T_{xx}}\rangle^{[2]}&=\chi_{tt}\langle{T_{tt}}\rangle^{[1]}+\chi_{xx}\langle{T_{xx}}\rangle^{[1]}+2\chi_{tx}\langle{T_{tx}}\rangle^{[1]}+\frac{1}{32\pi G}\Big[\partial_{x}\chi_{tt}\partial_{x}(\chi_{tt}+\chi_{xx})\notag\\
    &\quad-2\partial_t\chi_{tt}\partial_x\chi_{tx}-2\partial_t\chi_{tx}\partial_x\chi_{xx}+\partial_t\chi_{xx}\partial_t(\chi_{tt}+\chi_{xx})\notag\\
    &\quad+2(\chi_{tt}+\chi_{xx})(\partial_x^2\chi_{tt}-2\partial_t\partial_x\chi_{tx}+\partial_t^2\chi_{xx})\Big],\label{3-pt Tensionless Trace relation}\\
    \partial_t\langle{T_{tt}}\rangle^{[2]}+\partial_{x}\langle{T_{tx}}\rangle^{[2]}&=\frac{1}{2}(2\partial_t\chi_{tt}+2\chi_{tt}\partial_t+2\partial_x\chi_{tx}+2\chi_{tx}\partial_x-\partial_t\chi_{xx})\langle{T_{tt}}\rangle^{[1]}+\frac{1}{2}\partial_t\chi_{xx}\langle{T_{xx}}\rangle^{[1]}\notag\\
    &\quad-\frac{1}{2}(\partial_x\chi_{tt}-4\partial_t\chi_{tx}-2\chi_{tx}\partial_t-\partial_x\chi_{xx}-2\chi_{xx}\partial_x)\langle{T_{tx}}\rangle^{[1]},\label{3-pt Tensionless Conservation equation 1}\\
    \partial_x\langle{T_{xx}}\rangle^{[2]}+\partial_{t}\langle{T_{tx}}\rangle^{[2]}&=\frac{1}{2}\partial_x\chi_{tt}\langle{T_{tt}}\rangle^{[1]}-\frac{1}{2}(\partial_x\chi_{tt}-2\partial_t\chi_{tx}-2\chi_{tx}\partial_x-2\partial_x\chi_{xx}-2\chi_{xx}\partial_x)\langle{T_{xx}}\rangle^{[1]}\notag\\
    &\quad+\frac{1}{2}(\partial_t\chi_{tt}+2\chi_{tt}\partial_t+4\partial_x\chi_{tx}+2\chi_{tx}\partial_x-\partial_t\chi_{xx})\langle{T_{tx}}\rangle^{[1]}.\label{3-pt Tensionless Conservation equation 2}
\end{align}
Once again, by employing the Neumann boundary condition on the EOW brane, we obtain the modified profile $f^{[2]}$ and the boundary condition of $\langle{T_{tx}}\rangle^{[2]}$,
\begin{align}
    f^{[2]}(z,t)&=\frac{z^2}{4}[\chi_{tt}(\partial_x\chi_{tt}-2\partial_t\chi_{tx})-\chi_{tx}\partial_t\chi_{tt}+\chi_{xx}(\partial_x\chi_{tt}-2\partial_t\chi_{tx})](t,0)\notag\\
    &\quad+\frac{z^4}{16}[(\partial_x\chi_{tt}-2\partial_t\chi_{tx})(16\pi G\langle{T_{xx}}\rangle^{[1]}+\partial_x^2\chi_{tt}-2\partial_t\partial_x\chi_{tx}+\partial_t^2\chi_{xx})](t,0),\label{3-pt tensionless brane profile}\\
    \langle{T_{tx}(t,0)}\rangle^{[2]}&=[\chi_{tx}\langle{T_{tt}}\rangle^{[1]}](t,0)-\frac{1}{32\pi G}[2\chi_{tt}(\partial_t\partial_x\chi_{tt}-2\partial_t^2\chi_{tx})+\partial_x\chi_{tt}\partial_t(2\chi_{tt}+\chi_{xx})\notag\\
    &\quad-2\chi_{tx}\partial_t^2\chi_{tt}-2\partial_t\chi_{tx}\partial_t(3\chi_{tt}+\chi_{xx})](t,0).\label{Tensionless boundary condition of Ttx2}
\end{align}
Taking the second-order variation of (\ref{Tensionless boundary condition of Ttx2}), we derive the boundary condition of the three-point correlator $\frac{\delta^2\langle{T_{tx}(t,x)}\rangle^{[2]}}{\delta\chi_{ij}(t_1,x_1)\delta\chi_{kl}(t_2,x_2)}$. Note that this boundary condition only contains terms proportional to either $\delta(x_1)$ or $\delta(x_2)$. Assuming that the two insertion points $\boldsymbol{x}_1$ and $\boldsymbol{x}_2$ are not located on the boundary of the dual BCFT$_2$, then the boundary condition is simplified to
\begin{align}
	\frac{\delta^2\langle{T_{tx}(t,0)}\rangle^{[2]}}{\delta\chi_{ij}(t_1,x_1)\delta\chi_{kl}(t_2,x_2)}=0,\ \ \ \text{when }x_1>0\text{ and }x_2>0. \label{reduced Tensionless boundary condition of Ttx2}
\end{align}
This is consistent with the Cardy condition in BCFT$_2$ \cite{Cardy:1984bb, Cardy:2004hm}. Meanwhile, by combining (\ref{3-pt Tensionless Trace relation})(\ref{3-pt Tensionless Conservation equation 1})(\ref{3-pt Tensionless Conservation equation 2}), we find the Laplacian equation of $\langle{T_{tx}}\rangle^{[2]}$,
\begin{align}
    &\quad(\partial_{t}^2+\partial_{x}^2)\langle{T_{tx}}\rangle^{[2]}\notag\\
    &=(\partial_t\partial_x\chi_{tt}+\frac{3}{2}\partial_t\chi_{tt}\partial_x+\frac{3}{2}\partial_x\chi_{tt}\partial_t+\chi_{tt}\partial_t\partial_x+(\partial_x^2-\partial_t^2)\chi_{tx}+3\partial_x\chi_{tx}\partial_x-3\partial_t\chi_{tx}\partial_t\notag\\
    &\quad+\chi_{tx}(\partial_x^2-\partial_t^2)-\partial_t\partial_x\chi_{xx}-\frac{3}{2}\partial_t\chi_{xx}\partial_x-\frac{3}{2}\partial_x\chi_{xx}\partial_t-\chi_{xx}\partial_t\partial_x)\langle{T_{tt}}\rangle^{[1]}\notag\\
    &\quad+\frac{1}{2}((\partial_t^2-\partial_x^2)\chi_{tt}+4\partial_t\partial_x\chi_{tx}+(\partial_x^2-\partial_t^2)\chi_{xx})\langle{T_{tx}}\rangle^{[1]}+\frac{1}{32\pi G}\Big[\partial_x^3\chi_{tt}\partial_t\chi_{tt}-\frac{3}{2}\partial_t\partial_x(\partial_x\chi_{tt})^2\notag\\
    &\quad+2\partial_t^2(\partial_x\chi_{tt}\partial_x\chi_{tx}-\partial_x^2\chi_{tt}\chi_{tx})+2\partial_t^2\chi_{tt}\partial_x^2\chi_{tx}+2\partial_x^3\chi_{tt}\partial_x\chi_{tx}-\partial_x((\partial_t^2+4\partial_x^2)\chi_{tt}\partial_t\chi_{xx})\notag\\
    &\quad-\partial_x(2\partial_t\partial_x^2\chi_{tt}\chi_{xx}+\partial_t\partial_x\chi_{tt}\partial_x\chi_{xx})-2\partial_t\partial_x\chi_{tt}\partial_t^2\chi_{xx}-\partial_x\chi_{tt}\partial_t(\partial_t^2+\partial_x^2)\chi_{xx}+4\partial_t^2(\chi_{tx}\partial_t\partial_x\chi_{tx})\notag\\
    &\quad-4\partial_x\chi_{tx}\partial_t\partial_x^2\chi_{tx}+4\partial_t\partial_x(\partial_x\partial_x\chi_{tx}\chi_{xx})+2\partial_t^2(\partial_x\chi_{tx}\partial_x\chi_{xx}-\chi_{tx}\partial_t^2\chi_{xx})+4\partial_t\partial_x^2\chi_{tx}\partial_t\chi_{xx}\notag\\
    &\quad+2\partial_t(\partial_t\chi_{tx}\partial_x^2\chi_{xx})-\frac{5}{2}\partial_t\partial_x(\partial_t\chi_{xx})^2-2\partial_x(\chi_{xx}\partial_t^3\chi_{xx})-\partial_t\chi_{xx}\partial_t^2\partial_x\chi_{xx}\Big].
\end{align}
Solving this equation using the boundary condition (\ref{reduced Tensionless boundary condition of Ttx2}) and the regularity condition at $x=\infty$, we obtain $\langle{T_{tx}(\boldsymbol{x})T_{ij}(\boldsymbol{x}_1)T_{kl}(\boldsymbol{x}_2)}\rangle$ when $x_1>0$ and $x_2>0$,
\begin{align}
    &\quad\langle{T_{tx}(\boldsymbol{x})T_{tt}(\boldsymbol{x}_1)T_{tt}(\boldsymbol{x}_2)}\rangle\notag\\
    &=(F_1(\boldsymbol{x};\boldsymbol{x}_1)\partial_{x_1}+G_1(\boldsymbol{x};\boldsymbol{x}_1))\langle{T_{tt}(\boldsymbol{x}_1)T_{tt}(\boldsymbol{x}_2)}\rangle+(F_2(\boldsymbol{x};\boldsymbol{x}_1)\partial_{x_1}+G_2(\boldsymbol{x};\boldsymbol{x}_1))\langle{T_{tx}(\boldsymbol{x}_1)T_{tt}(\boldsymbol{x}_2)}\rangle\notag\\
      &\quad-\frac{1}{8\pi G}\Big[F_1(\boldsymbol{x};\boldsymbol{x}_1)\partial_{x_1}^3-\frac{3}{4}G_1(\boldsymbol{x};\boldsymbol{x}_1)\partial_{x_1}^2\Big]\delta(\boldsymbol{x}_1-\boldsymbol{x}_2)-\delta(\boldsymbol{x}-\boldsymbol{x}_1)\langle{T_{tx}(\boldsymbol{x}_1)T_{tt}(\boldsymbol{x}_2)}\rangle\notag\\
      &\quad+(\boldsymbol{x}_1\leftrightarrow \boldsymbol{x}_2),\\
      &\quad\langle{T_{tx}(\boldsymbol{x})T_{tt}(\boldsymbol{x}_1)T_{xx}(\boldsymbol{x}_2)}\rangle\notag\\
    &=(F_1(\boldsymbol{x};\boldsymbol{x}_1)\partial_{x_1}+G_1(\boldsymbol{x};\boldsymbol{x}_1))\langle{T_{tt}(\boldsymbol{x}_1)T_{xx}(\boldsymbol{x}_2)}\rangle+(F_2(\boldsymbol{x};\boldsymbol{x}_1)\partial_{x_1}+G_2(\boldsymbol{x};\boldsymbol{x}_1))\langle{T_{tx}(\boldsymbol{x}_1)T_{xx}(\boldsymbol{x}_2)}\rangle\notag\\
    &\quad-(F_1(\boldsymbol{x};\boldsymbol{x}_2)\partial_{x_2}+G_1(\boldsymbol{x};\boldsymbol{x}_2))\langle{T_{tt}(\boldsymbol{x}_1)T_{tt}(\boldsymbol{x}_2)}\rangle-(F_2(\boldsymbol{x};\boldsymbol{x}_2)\partial_{x_2}+G_2(\boldsymbol{x};\boldsymbol{x}_2))\langle{T_{tt}(\boldsymbol{x}_1)T_{tx}(\boldsymbol{x}_2)}\rangle\notag\\
    &\quad-\frac{1}{16\pi G}\Big[F_1(\boldsymbol{x};\boldsymbol{x}_1)\partial_{x_1}^3-F_2(\boldsymbol{x};\boldsymbol{x}_1)\partial_{t_1}^3+\frac{1}{2}G_1(\boldsymbol{x};\boldsymbol{x}_1)(2\partial_{t_1}^2+\partial_{x_1}^2)\notag\\
    &\quad+2F_1(\boldsymbol{x};\boldsymbol{x}_2)\partial_{x_2}^3-F_2(\boldsymbol{x};\boldsymbol{x}_2)(\partial_{t_2}^3+\partial_{t_2}\partial_{x_2}^2)+\frac{1}{2}G_1(\boldsymbol{x};\boldsymbol{x}_2)(\partial_{t_2}^2+4\partial_{x_2}^2)\Big]\delta(\boldsymbol{x}_1-\boldsymbol{x}_2)\notag\\
    &\quad-\delta(\boldsymbol{x}-\boldsymbol{x}_1)\langle{T_{tx}(\boldsymbol{x}_1)T_{xx}(\boldsymbol{x}_2)}\rangle+\delta(\boldsymbol{x}-\boldsymbol{x}_2)\langle{T_{tt}(\boldsymbol{x}_1)T_{tx}(\boldsymbol{x}_2)}\rangle,\\
      &\quad\langle{T_{tx}(\boldsymbol{x})T_{xx}(\boldsymbol{x}_1)T_{xx}(\boldsymbol{x}_2)}\rangle\notag\\
    &=-(F_1(\boldsymbol{x};\boldsymbol{x}_1)\partial_{x_1}+G_1(\boldsymbol{x};\boldsymbol{x}_1))\langle{T_{tt}(\boldsymbol{x}_1)T_{xx}(\boldsymbol{x}_2)}\rangle-(F_2(\boldsymbol{x};\boldsymbol{x}_1)\partial_{x_1}+G_2(\boldsymbol{x};\boldsymbol{x}_1))\langle{T_{tx}(\boldsymbol{x}_1)T_{xx}(\boldsymbol{x}_2)}\rangle\notag\\
      &\quad-\frac{1}{16\pi G}\Big[F_1(\boldsymbol{x};\boldsymbol{x}_1)\partial_{t_1}^2\partial_{x_1}+\frac{5}{2}G_1(\boldsymbol{x};\boldsymbol{x}_1)\partial_{t_1}^2-3F_2(\boldsymbol{x};\boldsymbol{x}_1)\partial_{t_1}^3\Big]\delta(\boldsymbol{x}_1-\boldsymbol{x}_2)\notag\\
      &\quad+\delta(\boldsymbol{x}-\boldsymbol{x}_1)\langle{T_{tx}(\boldsymbol{x}_1)T_{xx}(\boldsymbol{x}_2)}\rangle+(\boldsymbol{x}_1\leftrightarrow \boldsymbol{x}_2),\\
       &\quad\langle{T_{tx}(\boldsymbol{x})T_{tx}(\boldsymbol{x}_1)T_{tx}(\boldsymbol{x}_2)}\rangle\notag\\
    &=(F_1(\boldsymbol{x};\boldsymbol{x}_1)\partial_{x_1}+G_1(\boldsymbol{x};\boldsymbol{x}_1))\langle{T_{tx}(\boldsymbol{x}_1)T_{tx}(\boldsymbol{x}_2)}\rangle-(F_2(\boldsymbol{x};\boldsymbol{x}_1)\partial_{x_1}+G_2(\boldsymbol{x};\boldsymbol{x}_1))\langle{T_{tt}(\boldsymbol{x}_1)T_{tx}(\boldsymbol{x}_2)}\rangle\notag\\
      &\quad-\frac{1}{16\pi G}\Big[F_1(\boldsymbol{x};\boldsymbol{x}_1)\partial_{t_1}^2\partial_{x_1}-\frac{1}{2}G_1(\boldsymbol{x};\boldsymbol{x}_1)\partial_{t_1}^2+F_2(\boldsymbol{x};\boldsymbol{x}_1)(\partial_{t_1}^3-\partial_{t_1}\partial_{x_1}^2)\Big]\delta(\boldsymbol{x}_1-\boldsymbol{x}_2)\notag\\
      &\quad+\delta(\boldsymbol{x}-\boldsymbol{x}_1)\langle{T_{tt}(\boldsymbol{x}_1)T_{tx}(\boldsymbol{x}_2)}\rangle+(\boldsymbol{x}_1\leftrightarrow \boldsymbol{x}_2),\\
           &\quad\langle{T_{tx}(\boldsymbol{x})T_{tx}(\boldsymbol{x}_1)T_{tt}(\boldsymbol{x}_2)}\rangle\notag\\
    &=(F_1(\boldsymbol{x};\boldsymbol{x}_1)\partial_{x_1}+G_1(\boldsymbol{x};\boldsymbol{x}_1))\langle{T_{tx}(\boldsymbol{x}_1)T_{tt}(\boldsymbol{x}_2)}\rangle-(F_2(\boldsymbol{x};\boldsymbol{x}_1)\partial_{x_1}+G_2(\boldsymbol{x};\boldsymbol{x}_1))\langle{T_{tt}(\boldsymbol{x}_1)T_{tt}(\boldsymbol{x}_2)}\rangle\notag\\
    &\quad+(F_1(\boldsymbol{x};\boldsymbol{x}_2)\partial_{x_2}+G_1(\boldsymbol{x};\boldsymbol{x}_2))\langle{T_{tx}(\boldsymbol{x}_1)T_{tt}(\boldsymbol{x}_2)}\rangle+(F_2(\boldsymbol{x};\boldsymbol{x}_2)\partial_{x_2}+G_2(\boldsymbol{x};\boldsymbol{x}_2))\langle{T_{tx}(\boldsymbol{x}_1)T_{tx}(\boldsymbol{x}_2)}\rangle\notag\\
    &\quad+\frac{1}{16\pi G}\Big[2F_1(\boldsymbol{x};\boldsymbol{x}_1)\partial_{t_1}\partial_{x_1}^2+F_2(\boldsymbol{x};\boldsymbol{x}_1)(\partial_{t_1}^2\partial_{x_1}-\partial_{x_1}^3)+2F_1(\boldsymbol{x};\boldsymbol{x}_2)\partial_{t_2}\partial_{x_2}^2+G_1(\boldsymbol{x};\boldsymbol{x}_2)\partial_{t_2}\partial_{x_2}\notag\\
    &\quad+F_2(\boldsymbol{x};\boldsymbol{x}_2)\partial_{t_2}^2\partial_{x_2}+\frac{1}{2}G_2(\boldsymbol{x};\boldsymbol{x}_2)(\partial_{t_2}^2+\partial_{x_2}^2)-\delta(\boldsymbol{x}-\boldsymbol{x}_2)\partial_{t_2}^2\Big]\delta(\boldsymbol{x}_1-\boldsymbol{x}_2)\notag\\
    &\quad+\delta(\boldsymbol{x}-\boldsymbol{x}_1)\langle{T_{tt}(\boldsymbol{x}_1)T_{tt}(\boldsymbol{x}_2)}\rangle-\delta(\boldsymbol{x}-\boldsymbol{x}_2)\langle{T_{tx}(\boldsymbol{x}_1)T_{tx}(\boldsymbol{x}_2)}\rangle,\\
      &\quad\langle{T_{tx}(\boldsymbol{x})T_{tx}(\boldsymbol{x}_1)T_{xx}(\boldsymbol{x}_2)}\rangle\notag\\
    &=(F_1(\boldsymbol{x};\boldsymbol{x}_1)\partial_{x_1}+G_1(\boldsymbol{x};\boldsymbol{x}_1))\langle{T_{tx}(\boldsymbol{x}_1)T_{xx}(\boldsymbol{x}_2)}\rangle-(F_2(\boldsymbol{x};\boldsymbol{x}_1)\partial_{x_1}+G_2(\boldsymbol{x};\boldsymbol{x}_1))\langle{T_{tt}(\boldsymbol{x}_1)T_{xx}(\boldsymbol{x}_2)}\rangle\notag\\
    &\quad-(F_1(\boldsymbol{x};\boldsymbol{x}_2)\partial_{x_2}+G_1(\boldsymbol{x};\boldsymbol{x}_2))\langle{T_{tx}(\boldsymbol{x}_1)T_{tt}(\boldsymbol{x}_2)}\rangle-(F_2(\boldsymbol{x};\boldsymbol{x}_2)\partial_{x_2}+G_2(\boldsymbol{x};\boldsymbol{x}_2))\langle{T_{tx}(\boldsymbol{x}_1)T_{tx}(\boldsymbol{x}_2)}\rangle\notag\\
    &\quad+\frac{1}{16\pi G}\Big[\frac{1}{2}G_1(\boldsymbol{x};\boldsymbol{x}_1)\partial_{t_1}\partial_{x_1}-2F_2(\boldsymbol{x};\boldsymbol{x}_1)\partial_{t_1}^2\partial_{x_1}-\frac{1}{2}G_2(\boldsymbol{x};\boldsymbol{x}_1)\partial_{t_1}^2+F_1(\boldsymbol{x};\boldsymbol{x}_2)\partial_{t_2}\partial_{x_2}^2\notag\\
    &\quad+\frac{3}{2}G_1(\boldsymbol{x};\boldsymbol{x}_2)\partial_{t_2}\partial_{x_2}-2F_2(\boldsymbol{x};\boldsymbol{x}_2)\partial_{t_2}^2\partial_{x_2}-\frac{1}{2}G_2(\boldsymbol{x};\boldsymbol{x}_2)\partial_{t_2}^2+\delta(\boldsymbol{x}-\boldsymbol{x}_2)\partial_{t_2}^2\Big]\delta(\boldsymbol{x}_1-\boldsymbol{x}_2)\notag\\
    &\quad+\delta(\boldsymbol{x}-\boldsymbol{x}_1)\langle{T_{tt}(\boldsymbol{x}_1)T_{xx}(\boldsymbol{x}_2)}\rangle+\delta(\boldsymbol{x}-\boldsymbol{x}_1)\langle{T_{tx}(\boldsymbol{x}_1)T_{tx}(\boldsymbol{x}_2)}\rangle,
\end{align}
where the functions $F_1,F_2,G_1,G_2$ are defined by
\begin{align}
    F_1(\boldsymbol{x};\boldsymbol{x}')&=-\frac{t-t'}{2\pi((t-t')^2+(x-x')^2)}+\frac{t-t'}{2\pi((t-t')^2+(x+x')^2)},\notag\\
    F_2(\boldsymbol{x};\boldsymbol{x}')&=\frac{x-x'}{2\pi((t-t')^2+(x-x')^2)}+\frac{x+x'}{2\pi((t-t')^2+(x+x')^2)},\notag\\
    G_1(\boldsymbol{x};\boldsymbol{x}')&=\frac{-2(t-t')(x-x')}{\pi((t-t')^2+(x-x')^2)^2}+\frac{-2(t-t')(x+x')}{\pi((t-t')^2+(x+x')^2)^2},\notag\\
    G_2(\boldsymbol{x};\boldsymbol{x}')&=-\frac{(t-t')^2-(x-x')^2}{\pi((t-t')^2+(x-x')^2)^2}+\frac{(t-t')^2-(x+x')^2}{\pi((t-t')^2+(x+x')^2)^2}.
\end{align}
The other three-point correlators can be determined by equations (\ref{3-pt Tensionless Trace relation})(\ref{3-pt Tensionless Conservation equation 1})(\ref{3-pt Tensionless Conservation equation 2}),
\begin{align}
&\quad\langle{T_{tt}(\boldsymbol{x})T_{tt}(\boldsymbol{x}_1)T_{tt}(\boldsymbol{x}_2)}\rangle\notag\\
    &=(\tilde{F}_2(\boldsymbol{x};\boldsymbol{x}_1)\partial_{x_1}+\tilde{G}_2(\boldsymbol{x};\boldsymbol{x}_1))\langle{T_{tt}(\boldsymbol{x}_1)T_{tt}(\boldsymbol{x}_2)}\rangle-(\tilde{F}_1(\boldsymbol{x};\boldsymbol{x}_1)\partial_{x_1}+\tilde{G}_1(\boldsymbol{x};\boldsymbol{x}_1))\langle{T_{tx}(\boldsymbol{x}_1)T_{tt}(\boldsymbol{x}_2)}\rangle\notag\\
       &\quad-\frac{1}{16\pi G}\Big[2\tilde F_2(\boldsymbol{x};\boldsymbol{x}_1)\partial_{x_1}^3-\frac{3}{2}\tilde G_2(\boldsymbol{x};\boldsymbol{x}_1)\partial_{x_1}^2-3\delta(\boldsymbol{x}-\boldsymbol{x}_1)\partial_{x_1}^2\Big]\delta(\boldsymbol{x}_1-\boldsymbol{x}_2)\notag\\
       &\quad-\delta(\boldsymbol{x}-\boldsymbol{x}_1)\langle{T_{tt}(\boldsymbol{x}_1)T_{tt}(\boldsymbol{x}_2)}\rangle+(\boldsymbol{x}_1\leftrightarrow \boldsymbol{x}_2),\\
      &\quad\langle{T_{tt}(\boldsymbol{x})T_{tt}(\boldsymbol{x}_1)T_{xx}(\boldsymbol{x}_2)}\rangle\notag\\
    &=(\tilde{F}_2(\boldsymbol{x};\boldsymbol{x}_1)\partial_{x_1}+\tilde{G}_2(\boldsymbol{x};\boldsymbol{x}_1))\langle{T_{tt}(\boldsymbol{x}_1)T_{xx}(\boldsymbol{x}_2)}\rangle-(\tilde{F}_1(\boldsymbol{x};\boldsymbol{x}_1)\partial_{x_1}+\tilde{G}_1(\boldsymbol{x};\boldsymbol{x}_1))\langle{T_{tx}(\boldsymbol{x}_1)T_{xx}(\boldsymbol{x}_2)}\rangle\notag\\
      &\quad-(\tilde{F}_2(\boldsymbol{x};\boldsymbol{x}_2)\partial_{x_2}+\tilde{G}_2(\boldsymbol{x};\boldsymbol{x}_2))\langle{T_{tt}(\boldsymbol{x}_1)T_{tt}(\boldsymbol{x}_2)}\rangle+(\tilde{F}_1(\boldsymbol{x};\boldsymbol{x}_2)\partial_{x_2}+\tilde{G}_1(\boldsymbol{x};\boldsymbol{x}_2))\langle{T_{tt}(\boldsymbol{x}_1)T_{tx}(\boldsymbol{x}_2)}\rangle\notag\\
    &\quad-\frac{1}{16\pi G}\Big[\tilde F_2(\boldsymbol{x};\boldsymbol{x}_1)\partial_{x_1}^3+\tilde F_1(\boldsymbol{x};\boldsymbol{x}_1)\partial_{t_1}^3+\frac{1}{2}\tilde G_2(\boldsymbol{x};\boldsymbol{x}_1)(2\partial_{t_1}^2+\partial_{x_1}^2)+2\tilde F_2(\boldsymbol{x};\boldsymbol{x}_2)\partial_{x_2}^3\notag\\
    &\quad+\tilde F_1(\boldsymbol{x};\boldsymbol{x}_2)(\partial_{t_2}^3+\partial_{t_2}\partial_{x_2}^2)+\frac{1}{2}\tilde G_2(\boldsymbol{x};\boldsymbol{x}_2)(\partial_{t_2}^2+4\partial_{x_2}^2)-\delta(\boldsymbol{x}-\boldsymbol{x}_1)(\partial_{t_1}^2+\partial_{x_1}^2)\notag\\
    &\quad-3\delta(\boldsymbol{x}-\boldsymbol{x}_2)\partial_{x_2}^2\Big]\delta(\boldsymbol{x}_1-\boldsymbol{x}_2)-\delta(\boldsymbol{x}-\boldsymbol{x}_1)\langle{T_{tt}(\boldsymbol{x}_1)T_{xx}(\boldsymbol{x}_2)}\rangle\notag\\
    &\quad+\delta(\boldsymbol{x}-\boldsymbol{x}_2)\langle{T_{tt}(\boldsymbol{x}_1)T_{tt}(\boldsymbol{x}_2)}\rangle,\\
      &\quad\langle{T_{tt}(\boldsymbol{x})T_{xx}(\boldsymbol{x}_1)T_{xx}(\boldsymbol{x}_2)}\rangle\notag\\
    &=-(\tilde{F}_2(\boldsymbol{x};\boldsymbol{x}_1)\partial_{x_1}+\tilde{G}_2(\boldsymbol{x};\boldsymbol{x}_1))\langle{T_{tt}(\boldsymbol{x}_1)T_{xx}(\boldsymbol{x}_2)}\rangle+(\tilde{F}_1(\boldsymbol{x};\boldsymbol{x}_1)\partial_{x_1}+\tilde{G}_1(\boldsymbol{x};\boldsymbol{x}_1))\langle{T_{tx}(\boldsymbol{x}_1)T_{xx}(\boldsymbol{x}_2)}\rangle\notag\\
     &\quad-\frac{1}{16\pi G}\Big[\tilde F_2(\boldsymbol{x};\boldsymbol{x}_1)\partial_{t_1}^2\partial_{x_1}+\frac{5}{2}\tilde G_2(\boldsymbol{x};\boldsymbol{x}_1)\partial_{t_1}^2+3\tilde F_1(\boldsymbol{x};\boldsymbol{x}_1)\partial_{t_1}^3-2\delta(\boldsymbol{x}-\boldsymbol{x}_1)\partial_{t_1}^2\Big]\delta(\boldsymbol{x}_1-\boldsymbol{x}_2)\notag\\
      &\quad+\delta(\boldsymbol{x}-\boldsymbol{x}_1)\langle{T_{tt}(\boldsymbol{x}_1)T_{xx}(\boldsymbol{x}_2)}\rangle+(\boldsymbol{x}_1\leftrightarrow \boldsymbol{x}_2),\\
         &\quad\langle{T_{xx}(\boldsymbol{x})T_{xx}(\boldsymbol{x}_1)T_{xx}(\boldsymbol{x}_2)}\rangle\notag\\
    &=(\tilde{F}_2(\boldsymbol{x};\boldsymbol{x}_1)\partial_{x_1}+\tilde{G}_2(\boldsymbol{x};\boldsymbol{x}_1))\langle{T_{tt}(\boldsymbol{x}_1)T_{xx}(\boldsymbol{x}_2)}\rangle-(\tilde{F}_1(\boldsymbol{x};\boldsymbol{x}_1)\partial_{x_1}+\tilde{G}_1(\boldsymbol{x};\boldsymbol{x}_1))\langle{T_{tx}(\boldsymbol{x}_1)T_{xx}(\boldsymbol{x}_2)}\rangle\notag\\
 &\quad+\frac{1}{16\pi G}\Big[\tilde F_2(\boldsymbol{x};\boldsymbol{x}_1)\partial_{t_1}^2\partial_{x_1}+\frac{5}{2}\tilde G_2(\boldsymbol{x};\boldsymbol{x}_1)\partial_{t_1}^2+3\tilde F_1(\boldsymbol{x};\boldsymbol{x}_1)\partial_{t_1}^3-2\delta(\boldsymbol{x}-\boldsymbol{x}_1)\partial_{t_1}^2\Big]\delta(\boldsymbol{x}_1-\boldsymbol{x}_2)\notag\\
      &\quad-3\delta(\boldsymbol{x}-\boldsymbol{x}_1)\langle{T_{tt}(\boldsymbol{x}_1)T_{xx}(\boldsymbol{x}_2)}\rangle+\frac{1}{8\pi G}\partial_t\delta(\boldsymbol{x}-\boldsymbol{x}_1)\partial_t\delta(\boldsymbol{x}-\boldsymbol{x}_2)+(\boldsymbol{x}_1\leftrightarrow \boldsymbol{x}_2).
\end{align}
where the functions $\tilde{F}_1,\tilde{F}_2,\tilde{G}_1,\tilde{G}_2$ are defined by
\begin{align}
    \tilde{F}_1(\boldsymbol{x};\boldsymbol{x}')&=-\frac{t-t'}{2\pi((t-t')^2+(x-x')^2)}-\frac{t-t'}{2\pi((t-t')^2+(x+x')^2)},\notag\\
    \tilde{F}_2(\boldsymbol{x};\boldsymbol{x}')&=\frac{x-x'}{2\pi((t-t')^2+(x-x')^2)}-\frac{x+x'}{2\pi((t-t')^2+(x+x')^2)},\notag\\
    \tilde{G}_1(\boldsymbol{x};\boldsymbol{x}')&=\frac{-2(t-t')(x-x')}{\pi((t-t')^2+(x-x')^2)^2}-\frac{-2(t-t')(x+x')}{\pi((t-t')^2+(x+x')^2)^2},\notag\\
    \tilde{G}_2(\boldsymbol{x};\boldsymbol{x}')&=-\frac{(t-t')^2-(x-x')^2}{\pi((t-t')^2+(x-x')^2)^2}-\frac{(t-t')^2-(x+x')^2}{\pi((t-t')^2+(x+x')^2)}.
\end{align}
\subsubsection{Recurrence relation}
In principle, with increasingly tedious calculations, one can obtain any higher-point correlator by utilizing the method in the previous subsection. We derive recurrence relations for a special class of correlators to express the higher-point correlators in terms of the lower-point ones. To illustrate this, we work with complex coordinates $(w,\bar w)=(t+ix,t-ix)$ and turn on the $\bar w\bar w$ component of the metric variation,
\begin{align}
    \text{d}s^2_{(0)}=\text{d}w\text{d}\bar w+\epsilon F(w,\bar w)\text{d}\bar w^2.
\end{align}
The stress tensor conservation law (\ref{conservation equation 1}) and trace relation (\ref{trace relation 1}) take the forms
\begin{align}
    \partial_{\bar w}\langle{T_{ww}}\rangle&=-\partial_{w}\langle{T_{w\bar w}}\rangle+\epsilon(3\partial_{w}F+2F\partial_{w})\langle{T_{ww}}\rangle,\label{n-th conservation law 1}\\
    \partial_{w}\langle{T_{\bar w\bar w}}\rangle&=-\partial_{\bar w}\langle{T_{w\bar w}}\rangle+\epsilon\partial_{\bar w}F\langle{T_{ww}}\rangle+2\epsilon(\partial_{w}F+F\partial_{w})\langle{T_{w\bar w}}\rangle,\label{n-th conservation law 2}\\
    \langle{T_{w\bar w}}\rangle&=\epsilon F\langle{T_{ww}}\rangle+\frac{\epsilon\partial^2_{w}F}{16\pi G}.\label{n-th trace relation}
\end{align}
By expanding these equations to $n$-th order\footnote{For simplicity, here we assume that $n\geq 3$, which corresponds to correlators involving four or more stress tensor insertions.} in $\epsilon$, we find that
\begin{align}
    &\quad\partial_{w}\partial_{\bar w}\Big(\langle{T_{ww}}\rangle^{[n]}-\langle{T_{\bar w\bar w}}\rangle^{[n]}+2F\langle{T_{w\bar w}}\rangle^{[n-1]}\Big)\notag\\
    &=\Big(2\partial_{w}^2F+3\partial_{w}F\partial_{w}+F\partial_{w}^2+\partial_{\bar w}F\partial_{\bar w}+F\partial_{\bar w}^2\Big)\langle{T_{ww}}\rangle^{[n-1]}.\label{n-th order laplacian equation}
\end{align}
Meanwhile, from the Neumann boundary condition (\ref{Neumann boundary condition BCFT2}), we can read off the variation of the brane profile
\begin{align}
    f^{[n]}(z,w)&=\frac{i}{4}(-1)^{n}[F^{n-1}(\partial_w+\partial_{\bar w})F](w,w)z^2+O(z^4),
\end{align}
and the boundary condition for $(\langle{T_{ww}}\rangle-\langle{T_{\bar w\bar w}}\rangle)$,
\begin{align}
    &\quad[\langle{T_{ww}}\rangle^{[n]}-\langle{T_{\bar w\bar w}}\rangle^{[n]}](w,w)\notag\\
    &=-2[F\langle{T_{ww}}\rangle^{[n-1]}+F\langle{T_{w\bar w}}\rangle^{[n-1]}](w,w)-\frac{(-1)^n}{16\pi G}[F^{n-2}(F(\partial_{w}+\partial_{\bar w})^2F\notag\\
    &\quad+\partial_{w}F(\partial_{w}+\partial_{\bar w})F+\frac{3(n-1)}{2}((\partial_{w}+\partial_{\bar w})F)^2)](w,w). \label{n-th order boundary condition}
\end{align}
To simplify the results, we assume that the insertion points $\boldsymbol{w}_a$ for $a=1,2,...,n$ are not located on the boundary of BCFT$_2$. Then, the n-th order variation of (\ref{n-th order boundary condition}) becomes the Cardy condition,
\begin{align}
	\frac{\delta^n[\langle{T_{ww}}\rangle^{[n]}-\langle{T_{\bar w\bar w}}\rangle^{[n]}](w,w)}{\prod_{i=1}^{n}\delta F(w_i,\bar w_i)}=0,\ \ \ \text{when }\text{Im}(w_a)>0\text{ for }a=1,2,...,n.\label{simplified n-th order boundary condition}
\end{align} 
Taking the $n$-th order variation of (\ref{n-th order laplacian equation}), and solving the Laplacian equation with (\ref{simplified n-th order boundary condition}) and the regularity at infinity, we obtain
\begin{align}
    &\quad\frac{\delta^n[\langle{T_{ww}}\rangle^{[n]}-\langle{T_{\bar w\bar w}}\rangle^{[n]}+2F\langle{T_{w\bar w}}\rangle^{[n-1]}](\boldsymbol{w})}{\prod_{i=1}^{n}\delta F(\boldsymbol{w}_i)}\notag\\
    &=-\frac{2}{\pi}\sum_{i=1}^{n}\Bigg\lbrace\Big[(\frac{1}{(w_i-w)^2}-\frac{1}{(w_i-\bar w)^2})-\frac{1}{2}(\frac{1}{w_i-w}-\frac{1}{w_i-\bar w})\partial_{w_i}\notag\\
    &\quad-\frac{1}{2}(\frac{1}{\bar w_i-w}-\frac{1}{\bar w_i-\bar w})\partial_{\bar w_i}\Big]\frac{\delta^{n-1}\langle{T_{ww}(\boldsymbol{w}_i)}\rangle^{[n-1]}}{\prod_{j\neq i}\delta F(\boldsymbol{w}_j)}\Bigg\rbrace.
\end{align}
Combining this solution with (\ref{n-th conservation law 1})(\ref{n-th conservation law 2})(\ref{n-th trace relation}), we find
\begin{align}
    \frac{\delta^n\langle{T_{ww}(\boldsymbol{w})}\rangle^{[n]}}{\prod_{i=1}^{n}\delta F(\boldsymbol{w}_i)}&=-\frac{2}{\pi}\sum_{i=1}^{n}\Bigg\lbrace\Big[\frac{1}{(w_i-w)^2}-\frac{1}{2}\frac{1}{w_i-w}\partial_{w_i}-\frac{1}{2}\frac{1}{\bar w_i-w}\partial_{\bar w_i}\Big]\frac{\delta^{n-1}\langle{T_{ww}(\boldsymbol{w}_i)}\rangle^{[n-1]}}{\prod_{j\neq i}\delta F(\boldsymbol{w}_j)}\Bigg\rbrace,\notag\\
   \frac{\delta^n\langle{T_{\bar w\bar w}(\boldsymbol{w})}\rangle^{[n]}}{\prod_{i=1}^{n}\delta F(\boldsymbol{w}_i)}&=-\frac{2}{\pi}\sum_{i=1}^{n}\Bigg\lbrace\Big[\frac{1}{(w_i-\bar w)^2}-\frac{1}{2}\frac{1}{w_i-\bar w}\partial_{w_i}-\frac{1}{2}\frac{1}{\bar w_i-\bar w}\partial_{\bar w_i}\Big]\frac{\delta^{n-1}\langle{T_{ww}(\boldsymbol{w}_i)}\rangle^{[n-1]}}{\prod_{j\neq i}\delta F(\boldsymbol{w}_j)}\notag\\
    &\quad-\pi\delta(\boldsymbol{w}-\boldsymbol{w}_i)\frac{\delta^{n-1}\langle{T_{w\bar w}(\boldsymbol{w}_i)}\rangle^{[n-1]}}{\prod_{j\neq i}\delta F(\boldsymbol{w}_j)}\Bigg\rbrace,\notag\\
    \frac{\delta^n\langle{T_{w\bar w}(\boldsymbol{w})}\rangle^{[n]}}{\prod_{i=1}^{n}\delta F(\boldsymbol{w}_i)}&=\sum_{i=1}^{n}\delta(\boldsymbol{w}-\boldsymbol{w}_i)\frac{\delta^{n-1}\langle{T_{ww}(\boldsymbol{w}_i)}\rangle^{[n-1]}}{\prod_{j\neq i}\delta F(\boldsymbol{w}_j)}.
\end{align}
Finally, we use the conservation law (\ref{n-th conservation law 1}) and the definition (\ref{holographic correlator}) to derive the recurrence relations
\begin{align}
    &\quad\langle{T(\boldsymbol{w})T(\boldsymbol{w}_1)...T(\boldsymbol{w}_n)}\rangle\notag\\
  &=\sum_{i=1}^{n}\bigg\lbrace\Big[\frac{2}{(w_i-w)^2}-\frac{1}{w_i-w}\partial_{w_i}\Big]\langle{T(\boldsymbol{w}_1)...T(\boldsymbol{w}_n)}\rangle\notag\\
    &\quad+\frac{\pi}{\bar w_i-w}\sum_{j\neq i}(2\partial_{w_i}\delta(\boldsymbol{w}_i-\boldsymbol{w}_j)-\delta(\boldsymbol{w}_i-\boldsymbol{w}_j)\partial_{w_j})\langle{T(\boldsymbol{w}_1)...T(\boldsymbol{w}_{i-1})T(\boldsymbol{w}_{i+1})...T(\boldsymbol{w}_n)}\rangle\bigg\rbrace,\\
&\quad\langle{\bar T(\boldsymbol{w})T(\boldsymbol{w}_1)...T(\boldsymbol{w}_n)}\rangle\notag\\
  &=\sum_{i=1}^{n}\bigg\lbrace\Big[\frac{2}{(w_i-\bar w)^2}-\frac{1}{w_i-\bar w}\partial_{w_i}\Big]\langle{T(\boldsymbol{w}_1)...T(\boldsymbol{w}_n)}\rangle\notag\\
    &\quad+\frac{\pi}{\bar w_i-\bar w}\sum_{j\neq i}(2\partial_{w_i}\delta(\boldsymbol{w}_i-\boldsymbol{w}_j)-\delta(\boldsymbol{w}_i-\boldsymbol{w}_j)\partial_{w_j})\langle{T(\boldsymbol{w}_1)...T(\boldsymbol{w}_{i-1})T(\boldsymbol{w}_{i+1})...T(\boldsymbol{w}_n)}\rangle\notag\\
    &\quad-2\pi\delta(\boldsymbol{w}-\boldsymbol{w}_i)\langle{T(\boldsymbol{w}_1)...T(\boldsymbol{w}_{i-1})\Theta(\boldsymbol{w}_{i})T(\boldsymbol{w}_{i+1})...T(\boldsymbol{w}_n)}\rangle\bigg\rbrace,\\
    &\quad\langle{\Theta(\boldsymbol{w})T(\boldsymbol{w}_1)...T(\boldsymbol{w}_n)}\rangle\notag\\
  &=-\sum_{i=1}^{n}\pi\delta(\boldsymbol{w}-\boldsymbol{w}_i)\langle{T(\boldsymbol{w}_1)...T(\boldsymbol{w}_n)}\rangle,
\end{align}
where we used the notation $(T,\bar T,\Theta)=(-2\pi T_{ww},-2\pi T_{w\bar w},-2\pi T_{\bar w\bar w})$. If we exclude the contact terms, these results align with the Ward identity in BCFT$_2$ \cite{Cardy:1984bb}.

\subsection{Correlators with non-zero tension}\label{subsection 2.3}
Now, let us consider the non-zero tension case. When performing the calculations in the previous section, we find that the Neumann boundary condition (\ref{2-pt Tensionless NBC}) becomes complicated, making it difficult to obtain an exact form of the deformed brane profile $f^{[1]}(z,t)$. However, for computing the two-point correlators, only the boundary value of $\langle{T_{tx}}\rangle^{[1]}$ at $x=0$ is required. This boundary condition can be determined from the Neumann boundary condition at the leading order in $z$. One can verify that the boundary condition of $\langle{T_{tx}}\rangle^{[1]}$ is always independent of the brane tension, indicating that the two-point correlators with non-zero tension are equivalent to those in the tensionless case.\par
This subsection provides an alternative approach to compute stress tensor correlators with a general tension. The basic idea is to switch to a suitable Fefferman-Graham coordinate system, in which the EOW brane is a constant radial coordinate surface. For Poincare AdS$_3$ background (\ref{Poincare AdS3 metric}), we use the coordinate transformation
\begin{align}
    z=\frac{\xi}{\cosh{\rho}},\ \ \ t=\tau,\ \ \ x=\xi\tanh{\rho}.\label{Poincare AdS to hyperbolic slice 1}
\end{align}
The new bulk metric is represented in terms of the hyperbolic slice of AdS$_2$,
\begin{align}
    \text{d}s^2=\text{d}\rho^2+\cosh^2{\rho}\Big[\frac{\text{d}\tau^2+\text{d}\xi^2}{\xi^2}\Big].
\end{align}
In the hyperbolic slicing coordinates, the EOW brane is located at $\rho=\rho^*$, which is related to the brane tension via
\begin{align}
    \sinh{\rho^*}=k=-\frac{T}{\sqrt{1-T^2}}.
\end{align}
\begin{figure}[H]
    \centering
\includegraphics[scale=1.40]{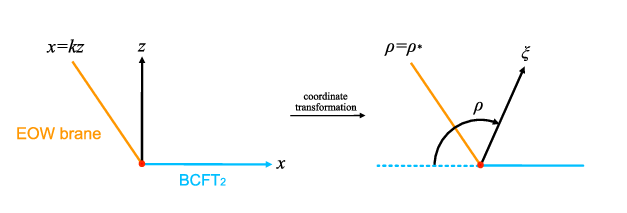}
    \caption{A sketch of Poincare AdS$_3$ coordinates (left) and hyperbolic slicing coordinates (right).}
\end{figure}\par
After the boundary metric perturbation (\ref{metric perturbation}), the coordinate transformation (\ref{Poincare AdS to hyperbolic slice 1}) should be modified accordingly,
\begin{align}
    z(\epsilon;\rho,\tau,\xi)&=\frac{\xi}{\cosh\rho}+\sum_{n=1}^{\infty}\epsilon^nz^{[n]}(\rho,\tau,\xi),\notag\\
    t(\epsilon;\rho,\tau,\xi)&=\tau+\sum_{n=1}^{\infty}\epsilon^nt^{[n]}(\rho,\tau,\xi),\notag\\
    x(\epsilon;\rho,\tau,\xi)&=\xi\tanh\rho+\sum_{n=1}^{\infty}\epsilon^nx^{[n]}(\rho,\tau,\xi),\label{Poincare AdS to hyperbolic slice 2}
\end{align}
The bulk metric in the (modified) hyperbolic slicing coordinates is
\begin{align}
    \mathcal{G}^{\text{H}}_{\mu\nu}=\frac{\partial x^a}{\partial \xi^{\mu}}\frac{\partial x^b}{\partial \xi^{\nu}}\mathcal{G}^{\text{P}}_{ab}. \label{modified bulk metric}
\end{align}
In the Fefferman-Graham gauge, the metric satisfies
\begin{align}
    \mathcal{G}^{\text{H}}_{\rho\rho}=1,\ \ \ \mathcal{G}^{\text{H}}_{\rho\tau}=\mathcal{G}^{\text{H}}_{\rho \xi}=0,\label{constrain from FG gauge}
\end{align}
where $\rho$ is related to the Fefferman-Graham radial coordinate by $z_{\text{FG}}=e^{-\rho}$. Plugging (\ref{metric perturbation})(\ref{bulk metric perturbation})(\ref{Poincare AdS to hyperbolic slice 2}) into (\ref{modified bulk metric}) and use the constrains (\ref{constrain from FG gauge}), we can determine the coefficients $z^{[n]}$, $t^{[n]}$, and $x^{[n]}$. At the first order in $\epsilon$, the constraints (\ref{constrain from FG gauge}) take the forms
\begin{align}
    &-\frac{(1+e^{-2\rho})z^{[1]}}{\xi e^{-3\rho}}-\frac{-2\partial_{\rho}x^{[1]}+e^{\rho}(1-e^{-2\rho})\partial_{\rho}z^{[1]}}{\xi e^{-2\rho}}\notag\\
    &\quad\quad\quad\quad\quad\quad\ +\frac{4(1+e^{-2\rho})^2\chi_{xx}(t^{[0]},x^{[0]})+16\xi^2e^{-2\rho}g^{[1]}_{(2)xx}(t^{[0]},x^{[0]})}{(1+e^{-2\rho})^4}=0,\label{FG gauge constrain 1}\\
    &-\frac{4\xi e^{-\rho}\partial_{\tau}x^{[1]}+e^{\rho}(1+e^{-2\rho})^2\partial_{\rho}t^{[1]}-2\xi(1-e^{-2\rho})\partial_{\tau}z^{[1]}}{4\xi^2e^{-2\rho}}\notag\\
    &\quad\quad\quad\quad\quad\quad\ -\frac{\chi_{tx}(t^{[0]},x^{[0]})}{\xi e^{-\rho}}-\frac{4\xi e^{-\rho}g^{[1]}_{(2)tx}(t^{[0]},x^{[0]})}{(1+e^{-2\rho})^2}=0,\label{FG gauge constrain 2}\\
    &-\frac{4\xi e^{-\rho}\partial_{\xi}x^{[1]}-2\xi(1-e^{-2\rho})\partial_{\xi}z^{[1]}-(1+e^{-2\rho})[-e^{\rho}(1-e^{-2\rho})\partial_{\rho}x^{[1]}-2\partial_{\rho}z^{[1]}]}{4\xi^2e^{-2\rho}}\notag\\
    &\quad\quad\quad\quad\quad\quad-\frac{(1-e^{-2\rho})\chi_{xx}(t^{[0]},x^{[0]})}{\xi e^{-\rho}(1+e^{-2\rho})}-\frac{4\xi e^{-\rho}(1-e^{-2\rho})g^{[1]}_{(2)xx}(t^{[0]},x^{[0]})}{(1+e^{-2\rho})^3}=0.\label{FG gauge constrain 3}
\end{align}
The next step is to expand the vector $x^{a[1]}$ in powers of $e^{-\rho}$,
\begin{align}
    x^{a[1]}(\rho,\tau,\xi)=\sum^{\infty}_{n=0}e^{-n\rho}x_{(n)}^{a[1]}(\tau,\xi). \label{FG expansion of coordinate transformation}
\end{align}
Plugging (\ref{FG expansion of coordinate transformation}) into (\ref{FG gauge constrain 1})(\ref{FG gauge constrain 2})(\ref{FG gauge constrain 3}), we can calculate the coefficients $x^{a[1]}_{(n)}$ order by order. Note that there are still three coefficients $t^{[1]}_{(0)}$, $x^{[1]}_{(0)}$, and $z^{[1]}_{(1)}$ that the above three equations cannot determine. By setting $t^{[1]}_{(0)}=x^{[1]}_{(0)}=0$, the coordinates on the conformal boundary in the hyperbolic slicing background can be aligned with those in the Poincare AdS$_3$ background,
\begin{align}
    \tau=t(\epsilon;\rho,\tau,\xi)\Big|_{\rho\to \infty},\ \ \ \xi=x(\epsilon;\rho,\tau,\xi)\Big|_{\rho\to \infty}.
\end{align}
Furthermore, we can set $z^{[1]}_{(1)}=0$ to ensure that the induced metric on a radial slice has the form of AdS$_2$. By imposing these three conditions, we obtain
\begin{align}
    z^{[1]}(\rho,\tau,\xi)&=2\xi e^{-3\rho}\chi_{xx}(\tau,\xi)+O(e^{-4\rho}),\notag\\
    t^{[1]}(\rho,\tau,\xi)&=2\xi e^{-2\rho}\chi_{tx}(\tau,\xi)+O(e^{-4\rho}),\notag\\
    x^{[1]}(\rho,\tau,\xi)&=2\xi e^{-2\rho}\chi_{xx}(\tau,\xi)+O(e^{-4\rho}).
\end{align}
Then, we can read off the Fefferman-Graham coefficients of the perturbed bulk metric in the hyperbolic slicing background,
\begin{align}
g^{[1]}_{(0)\alpha\beta}d\xi^{\alpha}d\xi^{\beta}&=\frac{\chi_{tt}(\tau,\xi)d\tau^2+2\chi_{tx}(\tau,\xi)d\tau d\xi+\chi_{xx}(\tau,\xi)d\xi^2}{4\xi^2},\notag\\
g^{[1]}_{(2)\alpha\beta}d\xi^{\alpha}d\xi^{\beta}&=\Big[g^{[1]}_{(2)tt}(\tau,\xi)+\frac{[(1-\xi\partial_\xi)\chi_{tt}+2\xi\partial_{\tau}\chi_{tx}-\chi_{xx}](\tau,\xi)}{2\xi^2}\Big]d\tau^2\notag\\
&\quad+2\Big[g^{[1]}_{(2)tx}(\tau,\xi)+\frac{[\chi_{tx}+\xi\partial_{\tau}\chi_{xx}](\tau,\xi)}{2\xi^2}\Big]d\tau d\xi\notag\\
&\quad+\Big[g^{[1]}_{(2)xx}(\tau,\xi)+\frac{\xi\partial_\xi\chi_{xx}(\tau,\xi)}{2\xi^2}\Big]d\xi^2. \label{Transformations of the FG coefficients}
\end{align}
The bulk Einstein's equation at the first-order in $\epsilon$ gives
\begin{align}
g^{[1]}_{(2)\tau\tau}+g^{[1]}_{(2)\xi\xi}&=2(1+3\xi\partial_{\xi}+\xi^2\partial^2_{\xi})g^{[1]}_{(0)\tau\tau}-4\xi\partial_{\tau}(1+\xi\partial_{\xi})g^{[1]}_{(0)\tau\xi}-2(1+\xi\partial_{\xi}-\xi^2\partial^2_{\tau})g^{[1]}_{(0)\xi\xi},\notag\\
    \partial_{\xi}g^{[1]}_{(2)\tau\tau}-\partial_{\tau}g^{[1]}_{(2)\tau \xi}&=-2(2\partial_{\xi}+\xi\partial^2_{\xi})g^{[1]}_{(0)\tau\tau}+2\partial_{\tau}(1+2\xi\partial_{\xi})g^{[1]}_{(0)\tau\xi}+2(\partial_{\xi}-\xi\partial^2_{\tau})g^{[1]}_{(0)\xi\xi},\notag\\
    \partial_{\tau}g^{[1]}_{(2)\xi\xi}-\partial_{\xi} g^{[1]}_{(2)\tau \xi}&=-2(\partial_{\xi}g^{[1]}_{(0)\tau\xi}-\partial_{\tau}g^{[1]}_{(0)\xi\xi}).\label{1st EOM of g2}
\end{align}
Next, we employ the Neumann boundary condition on the EOW brane. As we perturb the boundary metric, the profile of the EOW brane changes, and this modified profile can be formally written as
\begin{align}
    Q:\ \ \ \rho(\tau,\xi)=\rho^*+\sum_{n=1}^{\infty}\epsilon^n\psi^{[n]}(\tau,\xi).\label{new profile}
\end{align}
By plugging (\ref{new profile}) into (\ref{Neumann boundary condition BCFT2}), and applying the first-order Einstein's equation (\ref{1st EOM of g2}), we obtain
\begin{align}
&\frac{(-1-\xi\partial_{\xi}+\xi^2\partial_{\tau}^2)\psi^{[1]}}{\xi^2}-2g^{[1]}_{(0)\tau\tau}+g^{[1]}_{(2)\tau\tau}=0,\notag\\
&\frac{(\partial_{\tau}+\xi\partial_{\tau}\partial_{\xi})\psi^{[1]}}{\xi}-2g^{[1]}_{(0)\tau\xi}+g^{[1]}_{(2)\tau \xi}=0,\notag\\
&\frac{(-1+\xi\partial_{\xi}+\xi^2\partial_{\xi}^2)\psi^{[1]}}{\xi^2}-2g^{[1]}_{(0)\xi\xi}+g^{[1]}_{(2)\xi\xi}=0.\label{1st NBC in hyperbolic slicing background}
\end{align}
The Neumann boundary condition in the hyperbolic slicing coordinates has a distinct interpretation compared to Poincare coordinates. In the Poincare coordinates, the Neumann boundary condition provides a differential equation for the brane profile $f^{[1]}$ and specifies the boundary value of $g^{[1]}_{(2)tx}$. However, in the hyperbolic slicing coordinates, it establishes a relationship between the bulk solution $g^{[1]}_{(2)\alpha\beta}$ and the brane profile $\psi^{[1]}$. By employing (\ref{Transformations of the FG coefficients}) together with the definition (\ref{stress tensor one-point correlators}), we can express the perturbed one-point correlators $\langle{T_{ij}}\rangle^{[1]}$ in terms of $\psi^{[1]}$,
\begin{align}
    \langle{T_{tt}}\rangle^{[1]}&=\frac{1}{16\pi G}\Big[-2(\partial^2_{t}-\frac{1}{x}\partial_{x}-\frac{1}{x^2})\psi^{[1]}+\frac{1}{x}(\partial_{x}-x\partial_{x}^2)\chi_{tt}\notag\\
    &\quad-\frac{2}{x}\partial_{t}(1-x\partial_{x})\chi_{tx}+\frac{1}{x^2}(1-x^2\partial_{t}^2)\chi_{xx}\Big],\notag\\
    \langle{T_{tx}}\rangle^{[1]}&=-\frac{1}{16\pi G}\Big[\frac{2}{x}\partial_{t}(1+x\partial_{x})\psi^{[1]}+\frac{1}{x}\partial_{t}\chi_{xx}\Big],\notag\\
    \langle{T_{xx}}\rangle^{[1]}&=\frac{1}{16\pi G}\Big[-2(\partial_{x}^2+\frac{1}{x}\partial_{x}-\frac{1}{x^2})\psi^{[1]}-\partial^2_{x}\chi_{tt}+2\partial_{t}\partial_{x}\chi_{tx}\notag\\
    &\quad+\frac{1}{x^2}(1-x\partial_{x}-x^2\partial_{t}^2)\chi_{xx}\Big],\label{determine T1 from NBC}
\end{align}
where we changed the notations of the variables $(\tau,\xi)\to(t,x)$. Therefore, our primary objective is to determine the brane profile, specifically its variation with respect to the boundary metric $\frac{\delta\psi^{[1]}}{\delta \chi_{ij}}$, and subsequently obtain all the two-point correlators from (\ref{determine T1 from NBC}). From the trace relation (\ref{2-pt Tensionless Trace relation}), we find
\begin{align}
    (\partial^2_{t}+\partial^2_{x}-\frac{2}{x^2})\frac{\delta \psi^{[1]}(\boldsymbol{x})}{\delta\chi_{ij}(\boldsymbol{x}_0)}&=\Big[\frac{1}{2x}\delta^{i}_{t}\delta^{j}_{t}(\partial_{x}-x\partial_{x}^2)-\frac{1}{2x}\delta^{i}_{t}\delta^{j}_{x}\partial_t(1-x\partial_{x})\notag\\
    &\quad+\frac{1}{x^2}\delta^{i}_{x}\delta^{j}_{x}(1-\frac{x}{2}\partial_{x}-\frac{x^2}{2}\partial_{t}^2)\Big]\delta(\boldsymbol{x}-\boldsymbol{x}_0), \label{EOM for profile}
\end{align}
Using the Fourier transformation $\frac{\delta\psi^{[1]}(\boldsymbol{x})}{\delta \chi_{ij}(\boldsymbol{x}_0)}=\int_{-\infty}^{\infty}d\omega e^{-i\omega t}x^{\frac{1}{2}}\varphi^{ij}(x;\boldsymbol{x}_0)$, we have
\begin{align}
    \Big[\partial_{x}^2+\frac{1}{x}\partial_{x}-(\omega^2+\frac{9}{4x^2})\Big]\varphi^{ij}&=x^{-\frac{1}{2}}\frac{e^{i\omega t_0}}{2\pi}\Big[\frac{1}{2x}\delta^{i}_{t}\delta^{j}_{t}(\partial_{x}-x\partial_{x}^2)-\frac{1}{2x}\delta^{i}_{t}\delta^{j}_{x}(-i\omega)(1-x\partial_{x})\notag\\
    &\quad+\frac{1}{x^2}\delta^{i}_{x}\delta^{j}_{x}(1-\frac{x}{2}\partial_{x}+\frac{\omega^2x^2}{2})\Big]\delta(x-x_0). \label{ODE of phi}
\end{align}
The general solution is
\begin{align}
    \varphi^{ij}(x;\boldsymbol{x}_0)=&A^{ij}(\boldsymbol{x}_0)\varphi_{1}(x)+B^{ij}(\boldsymbol{x}_0)\varphi_{2}(x)+\varphi^{ij}_{*}(x;\boldsymbol{x}_0), \label{general solution for phi}
\end{align}
Here $\varphi_1(x)$ and $\varphi_2(x)$ are two homogeneous solutions
\begin{align}
    \varphi_{1}(x)&=\frac{1}{(|\omega|x)^{\frac{1}{2}}}\Big[\frac{\sinh{(|\omega|x)}}{|\omega|x}-\cosh{(|\omega|x)}\Big],\label{homogeneous solution 1}\\
        \varphi_{2}(x)&=\frac{1}{(|\omega|x)^{\frac{1}{2}}}\Big[\sinh{(|\omega|x)}-\frac{\cosh{(|\omega|x)}}{|\omega|x}\Big], \label{homogeneous solution 2}
\end{align}
which satisfy
\begin{align}
    \Big[\partial_{x}^2+\frac{1}{x}\partial_{x}-(\omega^2+\frac{9}{4x^2})\Big]\varphi_1=\Big[\partial_{x}^2+\frac{1}{x}\partial_{x}-(\omega^2+\frac{9}{4x^2})\Big]\varphi_2=0.
\end{align}
The last term $\varphi^{ij}_{*}(x;\boldsymbol{x}_0)$ is a particular solution of (\ref{ODE of phi}),\begin{align}
        \varphi^{ij}_{*}(x;\boldsymbol{x}_0)&=-\delta^{i}_{t}\delta^{j}_{t}\frac{e^{i\omega t_0}}{8\pi x_0^{\frac{3}{2}}}\Big[H(x-x_0)(\varphi_{1}(x)\tilde\varphi_{2}(x_0)-\varphi_{2}(x)\tilde\varphi_{1}(x_0))+2x_0\delta(x-x_0)\Big]\notag\\
        &\quad+\delta^{i}_{t}\delta^{j}_{x}\frac{i\omega e^{i\omega t_0}}{8\pi x_0^{\frac{1}{2}}}H(x-x_0)\Big[\varphi_{1}(x)\hat\varphi_2(x_0)-\varphi_{2}(x)\hat\varphi_1(x_0)\Big]\notag\\
        &\quad+\delta^{i}_{x}\delta^{j}_{x}\frac{e^{i\omega t_0}}{8\pi x_0^{\frac{3}{2}}}H(x-x_0)\Big[\varphi_{1}(x)\tilde\varphi_{2}(x_0)-\varphi_{2}(x)\tilde\varphi_{1}(x_0)\Big],
\end{align}
where $\tilde{\varphi}(x)=(3+2\omega^2x^2+2x\partial_{x})\varphi(x)$ and $\hat{\varphi}(x)=(3+2x\partial_x)\varphi(x)$. $H(x-x_0)$ is the Heaviside step function defined as $H(x-x_0)=1$ for $x>x_0$, $H(x-x_0)=\frac{1}{2}$ for $x=x_0$, and $H(x-x_0)=0$ for $x<x_0$. The coefficients $A^{ij}$ and $B^{ij}$ are fixed by imposing the regularity conditions on $\psi^{[1]}$ at both $x=0$ and $x=\infty$,
\begin{align}
    A^{ij}(\boldsymbol{x}_0)&=(\delta^{i}_{t}\delta^{j}_{t}-\delta^{i}_{x}\delta^{j}_{x})\frac{e^{i\omega t_0}}{8\pi x_0^{\frac{3}{2}}}(\tilde\varphi_1(x_0)+\tilde\varphi_2(x_0))-\delta^{i}_{t}\delta^{j}_{x}\frac{i\omega e^{i\omega t_0}}{8\pi x_0^{\frac{1}{2}}}(\hat\varphi_1(x_0)+\hat\varphi_2(x_0)),\notag\\
    B^{ij}(\boldsymbol{x}_0)&=0.\label{constrains of 2-pt 1}
\end{align}
Putting everything together, we obtain the two-point correlators in frequency space,
\begin{align}
\langle{T_{tx}(\omega,x)T_{tt}(-\omega,x_0)}\rangle&=-\frac{i\omega}{16\pi^2G}\Big[\frac{\omega^2}{2}\Big(e^{-|\omega|(x+x_0)}+H(x-x_0)e^{-|\omega|(x-x_0)}\notag\\
&\quad-H(x_0-x)e^{|\omega|(x-x_0)}\Big)+\partial_{x}\delta(x-x_0)\Big],\notag\\
      \langle{T_{tx}(\omega,x)T_{tx}(-\omega,x_0)}\rangle&=-\frac{(i\omega)^2}{16\pi^2G}\Big[\frac{|\omega|}{2}\Big(e^{-|\omega|(x+x_0)}-H(x-x_0)e^{-|\omega|(x-x_0)}\notag\\
      &\quad-H(x_0-x)e^{|\omega|(x-x_0)}\Big)+\delta(x-x_0)\Big],\notag\\
 \langle{T_{tx}(\omega,x)T_{xx}(-\omega,x_0)}\rangle&=\frac{i\omega}{16\pi^2G}\Big[\frac{\omega^2}{2}\Big(e^{-|\omega|(x+x_0)}+H(x-x_0)e^{-|\omega|(x-x_0)}\notag\\
 &\quad-H(x_0-x)e^{|\omega|(x-x_0)}\Big)\Big],\notag\\
   \langle{T_{tt}(\omega,x)T_{tt}(-\omega,x_0)}\rangle&=\frac{1}{16\pi^2G}\Big[\frac{|\omega|^3}{2}\Big(e^{-|\omega|(x+x_0)}+e^{-|\omega(x-x_0)|}\Big)-(\omega^2+\partial_{x}^2)\delta(x-x_0)\Big],\notag\\
     \langle{T_{tt}(\omega,x)T_{xx}(-\omega,x_0)}\rangle&=-\frac{1}{16\pi^2G}\Big[\frac{|\omega|^3}{2}\Big(e^{-|\omega|(x+x_0)}+e^{-|\omega(x-x_0)|}\Big)-\omega^2\delta(x-x_0)\Big],\notag\\
  \langle{T_{xx}(\omega,x)T_{xx}(-\omega,x_0)}\rangle&=\frac{1}{16\pi^2G}\Big[\frac{|\omega|^3}{2}\Big(e^{-|\omega|(x+x_0)}+e^{-|\omega(x-x_0)|}\Big)\Big],
\end{align}
which match the results (\ref{tensionless 2-pt result 1})(\ref{tensionless 2-pt result 2}) obtained by the previous method. The first-order brane profile $\psi^{[1]}$ and the two-point correlators are independent of brane tension. Next, we will demonstrate that this property also holds for three-point correlators.\par
From above method, we find the coordinate transformation (\ref{Poincare AdS to hyperbolic slice 2}) at the second order in $\epsilon$,
\begin{align}
z^{[2]}(\rho,\tau,\xi)=&-2\xi e^{-3\rho}[\chi^2_{tx}+\chi^2_{xx}](\tau,\xi)+O(e^{-4\rho}),\notag\\
t^{[2]}(\rho,\tau,\xi)=&-2\xi e^{-2\rho}[\chi_{tt}\chi_{tx}+\chi_{tx}\chi_{xx}](\tau,\xi)+O(e^{-4\rho}),\notag\\
x^{[2]}(\rho,\tau,\xi)=&-2\xi e^{-2\rho}[\chi^2_{tx}+\chi^2_{xx}](\tau,\xi)+O(e^{-4\rho}).\label{second order coordinate transformation}
\end{align}
Plugging (\ref{second order coordinate transformation}) into (\ref{modified bulk metric}), we obtain the Fefferman-Graham coefficients,
\begin{align}
g^{[2]}_{(0)\alpha\beta}d\xi^{\alpha}d\xi^{\beta}&=0,\notag\\
g^{[2]}_{(2)\alpha\beta}d\xi^{\alpha}d\xi^{\beta}&=\Big[g^{[2]}_{(2)tt}(\tau,\xi)+\frac{[\chi_{tx}^2+\chi_{xx}^2-(1-\xi\partial_\xi)\chi_{tt}\chi_{xx}-2\xi\partial_{\tau}\chi_{tx}\chi_{xx}-\xi\partial_{\tau}\chi_{tt}\chi_{tx}](\tau,\xi)}{2\xi^2}\Big]d\tau^2\notag\\
&\quad+2\Big[g^{[2]}_{(2)tx}(\tau,\xi)-\frac{[\chi_{tx}\chi_{xx}+\xi\partial_{\xi}\chi_{tt}\chi_{tx}+\xi\partial_{\tau}\chi_{xx}\chi_{xx}](\tau,\xi)}{2\xi^2}\Big]d\tau d\xi\notag\\
&\quad+\Big[g^{[2]}_{(2)xx}(\tau,\xi)+\frac{[\chi_{tx}^2-2\xi\partial_\xi\chi_{tx}\chi_{tx}+\xi\partial_{\tau}\chi_{xx}\chi_{tx}-\xi\partial_{\xi}\chi_{xx}\chi_{xx}](\tau,\xi)}{2\xi^2}\Big]d\xi^2. \label{3-pt Transformations of the FG coefficients}
\end{align}
Expanding the Neumann boundary condition (\ref{Neumann boundary condition BCFT2}) to the second order in $\epsilon$ and using (\ref{stress tensor one-point correlators})(\ref{determine T1 from NBC})(\ref{EOM for profile}), we can express $\langle{T_{ij}}\rangle^{[2]}$ in terms of $\psi^{[1]}$ and $\psi^{[2]}$,
\begin{align}
    \langle{T_{tt}}\rangle^{[2]}&=\frac{1}{16\pi G}\Big[-2(\partial_{t}^2-\frac{1}{x}\partial_{x}-\frac{1}{x^2})\psi^{[2]}+\frac{2}{x^2}\psi^{[1]2}+(4+3T)(\partial_{t}\psi^{[1]})^2+(2+3T)(\partial_{x}\psi^{[1]})^2\notag\\
    &\quad-(1+T)(x^2\partial_{t}^2-x\partial_{x})[(\partial_{t}\psi^{[1]})^2+(\partial_{x}\psi^{[1]})^2]+\frac{2}{x^2}\chi_{tt}(1+x\partial_{x})\psi^{[1]}+\partial_{t}\chi_{tt}\partial_{t}\psi^{[1]}\notag\\
    &\quad-\partial_{x}\chi_{tt}\partial_{x}\psi^{[1]}-\frac{2}{x}\chi_{tx}\partial_{t}\psi^{[1]}+2\partial_{t}\chi_{tx}\partial_{x}\psi^{[1]}-\frac{2}{x}\chi_{xx}\partial_{x}\psi^{[1]}+\mathcal{F}^{[2]}_{tt}\Big],\notag\\
    \langle{T_{tx}}\rangle^{[2]}&=-\frac{1}{16\pi G}\Big[\frac{2}{x}\partial_{t}(1+x\partial_{x})\psi^{[2]}-2\partial_{t}\psi^{[1]}\partial_{x}\psi^{[1]}+(1+T)x\partial_{t}(3+x\partial_{x})[(\partial_{t}\psi^{[1]})^2+(\partial_{x}\psi^{[1]})^2]\notag\\
    &\quad-\frac{1}{x^2}\partial_{x}\chi_{tt}\partial_{t}\psi^{[1]}-\frac{2}{x^2}\chi_{tx}(1+x\partial_{x})\psi^{[1]}-\partial_{t}\chi_{xx}\partial_{x}\psi^{[1]}+\mathcal{F}^{[2]}_{tx}\Big],\notag\\
    \langle{T_{xx}}\rangle^{[2]}&=\frac{1}{16\pi G}\Big[-2(\partial_{x}^2+\frac{1}{x}\partial_{x}-\frac{1}{x^2})\psi^{[2]}+\frac{2}{x^2}\psi^{[1]2}-(4+3T)(\partial_{t}\psi^{[1]})^2-(2+3T)(\partial_{x}\psi^{[1]})^2\notag\\
    &\quad-(1+T)(x^2\partial_{x}^2+5x\partial_{x})[(\partial_{t}\psi^{[1]})^2+(\partial_{x}\psi^{[1]})^2]-\frac{2}{x}(1-x\partial_{x})\chi_{tx}\partial_{t}\psi^{[1]}\notag\\
    &\quad+\frac{2}{x^2}\chi_{xx}\psi^{[1]}+\partial_{x}\chi_{xx}\partial_{x}\psi^{[1]}-\partial_{t}\chi_{xx}\partial_{t}\psi^{[1]}+\mathcal{F}^{[2]}_{xx}\Big],\label{determine T2}
\end{align}
where $\mathcal{F}^{[2]}_{tt}$, $\mathcal{F}^{[2]}_{tx}$, and $\mathcal{F}^{[2]}_{xx}$ are quadratic functions of $\chi_{ij}$ that are independent of brane tension. The equation of motion for $\frac{\delta^2\psi^{[2]}}{\delta\chi_{i_1j_1}\delta\chi_{i_2j_2}}$ can be derived by taking functional derivative of (\ref{determine T2}) and employing the trace relation (\ref{3-pt Tensionless Trace relation}),
\begin{align}
    &\quad(\partial_{t}^2+\partial_{x}^2-\frac{2}{x^2})\frac{\delta^2\psi^{[2]}(\boldsymbol{x})}{\delta\chi_{i_1j_1}(\boldsymbol{x}_1)\delta\chi_{i_2j_2}(\boldsymbol{x}_2)}\notag\\
    &=\frac{2}{x^2}\frac{\delta\psi^{[1]}(\boldsymbol{x})}{\delta\chi_{i_1j_1}(\boldsymbol{x}_1)}\frac{\delta\psi^{[1]}(\boldsymbol{x})}{\delta\chi_{i_2j_2}(\boldsymbol{x}_2)}-\frac{x}{2}(1+T)(x\partial_{t}^2+x\partial_{x}^2+4\partial_{x})\Big[\partial_{t}\frac{\delta\psi^{[1]}(\boldsymbol{x})}{\delta\chi_{i_1j_1}(\boldsymbol{x}_1)}\partial_{t}\frac{\delta\psi^{[1]}(\boldsymbol{x})}{\delta\chi_{i_2j_2}(\boldsymbol{x}_2)}\notag\\
    &\quad+\partial_{x}\frac{\delta\psi^{[1]}(\boldsymbol{x})}{\delta\chi_{i_1j_1}(\boldsymbol{x}_1)}\partial_{x}\frac{\delta\psi^{[1]}(\boldsymbol{x})}{\delta\chi_{i_2j_2}(\boldsymbol{x}_2)}\Big]+\frac{1}{2}\Big[\delta^{i_1}_{t}\delta^{j_1}_{t}(\partial_{t}\delta(\boldsymbol{x}-\boldsymbol{x}_1)\partial_{t}-\partial_{x}\delta(\boldsymbol{x}-\boldsymbol{x}_1)\partial_{x}+2\delta(\boldsymbol{x}-\boldsymbol{x}_1)\partial_{t}^2)\notag\\
    &\quad+\delta^{i_1}_{t}\delta^{j_1}_{x}(\partial_{x}\delta(\boldsymbol{x}-\boldsymbol{x}_1)\partial_{t}+\partial_{t}\delta(\boldsymbol{x}-\boldsymbol{x}_1)\partial_{x}+2\delta(\boldsymbol{x}-\boldsymbol{x}_1)\partial_{t}\partial_{x})+\delta^{i_1}_{x}\delta^{j_1}_{x}(\partial_{x}\delta(\boldsymbol{x}-\boldsymbol{x}_1)\partial_{x}\notag\\
    &\quad-\partial_{t}\delta(\boldsymbol{x}-\boldsymbol{x}_1)\partial_{t}+2\delta(\boldsymbol{x}-\boldsymbol{x}_1)\partial_{x}^2)\Big]\frac{\delta\psi^{[1]}(\boldsymbol{x})}{\delta\chi_{i_2j_2}(\boldsymbol{x}_2)}+\frac{1}{4}\frac{\delta^2[\mathcal{F}^{[2]}_{tt}+\mathcal{F}^{[2]}_{xx}](\boldsymbol{x})}{\delta\chi_{i_1j_1}(\boldsymbol{x}_1)\delta\chi_{i_2j_2}(\boldsymbol{x}_2)}+(\text{1}\leftrightarrow\text{2}).\label{EOM for psi2}
\end{align}
We are concerned with the dependence of three-point correlators on the brane tension $T$. Let us divide the brane profile into two parts,
\begin{align}
    \psi^{[2]}(\boldsymbol{x})=\psi^{[2]}_{0}(\boldsymbol{x})+\Delta\psi^{[2]}(\boldsymbol{x}),\label{decomposed brane profile}
\end{align}
where $\psi^{[2]}_{0}(\boldsymbol{x})$ is the tensionless brane profile. Plugging (\ref{decomposed brane profile}) into (\ref{EOM for psi2}), and using the fact that $\frac{\delta\psi^{[1]}}{\delta\chi_{ij}}$ is independent of brane tension, we obtain
\begin{align}
   &(\partial_{t}^2+\partial_{x}^2-\frac{2}{x^2})\Big[\frac{\delta^2\Delta\psi^{[2]}(\boldsymbol{x})}{\delta\chi_{i_1j_1}(\boldsymbol{x}_1)\delta\chi_{i_2j_2}(\boldsymbol{x}_2)}\notag\\
   &\quad\quad\quad+x^2T\Big(\partial_{t}\frac{\delta\psi^{[1]}(\boldsymbol{x})}{\delta\chi_{i_1j_1}(\boldsymbol{x}_1)}\partial_{t}\frac{\delta\psi^{[1]}(\boldsymbol{x})}{\delta\chi_{i_2j_2}(\boldsymbol{x}_2)}+\partial_{x}\frac{\delta\psi^{[1]}(\boldsymbol{x})}{\delta\chi_{i_1j_1}(\boldsymbol{x}_1)}\partial_{x}\frac{\delta\psi^{[1]}(\boldsymbol{x})}{\delta\chi_{i_2j_2}(\boldsymbol{x}_2)}\Big)\Big]=0.
\end{align}
We impose the regularity conditions on $\psi^{[2]}$ at $x=0,\infty$ to get
\begin{align}
	\frac{\delta^2\psi^{[2]}(\boldsymbol{x})}{\delta\chi_{i_1j_1}(\boldsymbol{x}_1)\delta\chi_{i_2j_2}(\boldsymbol{x}_2)}&=\frac{\delta^2\psi_{0}^{[2]}(\boldsymbol{x})}{\delta\chi_{i_1j_1}(\boldsymbol{x}_1)\delta\chi_{i_2j_2}(\boldsymbol{x}_2)}-x^2T\Big(\partial_{t}\frac{\delta\psi^{[1]}(\boldsymbol{x})}{\delta\chi_{i_1j_1}(\boldsymbol{x}_1)}\partial_{t}\frac{\delta\psi^{[1]}(\boldsymbol{x})}{\delta\chi_{i_2j_2}(\boldsymbol{x}_2)}\notag\\
	&\quad+\partial_{x}\frac{\delta\psi^{[1]}(\boldsymbol{x})}{\delta\chi_{i_1j_1}(\boldsymbol{x}_1)}\partial_{x}\frac{\delta\psi^{[1]}(\boldsymbol{x})}{\delta\chi_{i_2j_2}(\boldsymbol{x}_2)}\Big).
\end{align}
Combining this with (\ref{determine T2}), we find
\begin{align}
    \frac{\delta^2\langle{T_{ij}(\boldsymbol{x})}\rangle}{\delta\chi_{i_1j_1}(\boldsymbol{x}_1)\delta\chi_{i_2j_2}(\boldsymbol{x}_2)}=\frac{\delta^2\langle{T_{ij}(\boldsymbol{x})}\rangle_{0}}{\delta\chi_{i_1j_1}(\boldsymbol{x}_1)\delta\chi_{i_2j_2}(\boldsymbol{x}_2)},
\end{align}
which indicates that three-point correlators are independent of brane tension.
\subsection{Other classical saddles}
At the end of this section, we consider other classical gravitational saddle points that are dual to excited states and calculate the stress tensor two-point correlators dominated by them. Three-dimensional gravity has no local degrees of freedom, and all classical saddle points can be constructed from Poincare AdS$_3$
\begin{align}
	\text{d}s^2=\frac{\text{d}Z^2+\text{d}W\text{d}\overline{W}}{Z^2}
\end{align}
via the Ba$\tilde{\text{n}}$ados map \cite{Banados:1998gg}
\begin{align}
	W&=p(w)-\frac{2z^2(\frac{\text{d}p}{\text{d}w})^2\frac{\text{d}^2\bar p}{\text{d}\bar w^2}}{4\frac{\text{d}p}{\text{d}w}\frac{\text{d}\bar p}{\text{d}\bar w}+z^2\frac{\text{d}^2p}{\text{d}w^2}\frac{\text{d}^2\bar p}{\text{d}\bar w^2}},\notag\\
	\overline{W}&=\bar p(\bar w)-\frac{2z^2(\frac{\text{d}\bar p}{\text{d}\bar w})^2\frac{\text{d}^2p}{\text{d}w^2}}{4\frac{\text{d}p}{\text{d}w}\frac{\text{d}\bar p}{\text{d}\bar w}+z^2\frac{\text{d}^2p}{\text{d}w^2}\frac{\text{d}^2\bar p}{\text{d}\bar w^2}},\notag\\
	Z&=\frac{4z(\frac{\text{d}p}{\text{d}w}\frac{\text{d}\bar p}{\text{d}\bar w})^{\frac{3}{2}}}{4\frac{\text{d}p}{\text{d}w}\frac{\text{d}\bar p}{\text{d}\bar w}+z^2\frac{\text{d}^2p}{\text{d}w^2}\frac{\text{d}^2\bar p}{\text{d}\bar w^2}}.\label{Banados map}
\end{align}
The bulk metric takes the form
\begin{align}
	\text{d}s^2&=\frac{\text{d}z^2}{z^2}+\frac{1}{z^2}\Big[\text{d}w\text{d}\bar w-z^2(\mathcal{T}(w)\text{d}w^2+\bar{\mathcal{T}}(\bar w)\text{d}\bar{w}^2)+z^4\mathcal{T}(w)\bar{\mathcal{T}}(\bar w)\text{d}w\text{d}\bar w\Big],\label{general saddle metric}
\end{align}
where
\begin{align}
	\mathcal{T}(w)=\frac{1}{2}\Big[\frac{\frac{\text{d}^3p}{\text{d}w^3}}{\frac{\text{d}p}{\text{d}w}}-\frac{3}{2}\frac{(\frac{\text{d}^2p}{\text{d}w^2})^2}{(\frac{\text{d}p}{\text{d}w})^2}\Big],\ \ \ \ \bar{\mathcal{T}}(\bar w)=\frac{1}{2}\Big[\frac{\frac{\text{d}^3\bar p}{\text{d}\bar w^3}}{\frac{\text{d}\bar p}{\text{d}\bar w}}-\frac{3}{2}\frac{(\frac{\text{d}^2\bar p}{\text{d}\bar w^2})^2}{(\frac{\text{d}\bar p}{\text{d}\bar w})^2}\Big].
\end{align}
The dual BCFT$_2$ is defined on the half plane $\text{Im}(W)\geq 0$ in the original coordinates. Following \cite{Izumi:2022opi}, we require that after the coordinate transformation (\ref{Banados map}), this right half plane is mapped to the same region (i.e., $\text{Im}(w)\geq 0$). This can be achieved by setting $p$ and $\bar p$ as the same function.\par
In the new background (\ref{general saddle metric}), the EOW brane profile can be obtained by imposing the Neumann boundary condition (\ref{Neumann boundary condition BCFT2}),
\begin{align}
	x(z,t)=-\frac{T}{\sqrt{1-T^2}}z+\frac{T(3-T^2)}{3(1-T^2)^{\frac{3}{2}}}\mathcal{T}(t)z^3+O(z^4),
\end{align}
where $t$ and $x$ represent the real and imaginary parts of $w$, respectively.\par 
As a simple example, we will consider the case where $\mathcal{T}$ is a constant and compute the stress tensor two-point correlators. Let us perturb the boundary metric as (\ref{metric perturbation}). The brane profile can be formally written as the power series (\ref{tensionless brane profile}) in $\epsilon$. By employing the Neumann boundary condition (\ref{Neumann boundary condition BCFT2}) at the first order in $\epsilon$, we obtain the modified brane profile $f^{[1]}$ and the boundary condition of $\langle{T_{tx}}\rangle^{[1]}$,
\begin{align}
	f^{[1]}(z,t)&=\frac{T\chi_{xx}(t,0)}{2\sqrt{1-T^2}}z-\frac{[\partial_x\chi_{tt}-2\partial_{t}\chi_{tx}+T^2\partial_x\chi_{xx}](t,0)}{4(1-T^2)}z^2+O(z^3),\notag\\
	\langle{T_{tx}(t,0)}\rangle^{[1]}&=\frac{1}{16\pi G}[\partial_{t}\partial_{x}\chi_{tt}-2\partial_{t}^2\chi_{tx}-4\mathcal{T}\chi_{tx}](t,0).
\end{align}
Meanwhile, $\langle{T_{tx}}\rangle^{[1]}$ satisfies the Laplacian equation
\begin{align}
	(\partial_t^2+\partial_x^2)\langle{T_{tx}}\rangle^{[1]}&=\frac{1}{16\pi G}(\partial_{t}\partial_{x}^3\chi_{tt}-2\partial_{t}^2\partial_{x}^2\chi_{tx}+\partial_t^3\partial_x\chi_{xx})\notag\\
	&\quad+\frac{\mathcal{T}}{4\pi G}(-\partial_t\partial_x\chi_{tt}+(\partial_t^2-\partial_x^2)\chi_{tx}+\partial_t\partial_x\chi_{xx}).
\end{align}
Putting everything together, we obtain the two-point correlators $\langle{T_{tx}T_{ij}}\rangle$,
\begin{align}
	\langle{T_{tx}(\boldsymbol{x})T_{tt}(\boldsymbol{x}_0)}\rangle&=\frac{3i}{16\pi^2 G}\Big[\frac{1}{(t-t_0+i(x+x_0))^4}-\frac{1}{(t-t_0-i(x+x_0))^4}\notag\\
	&\quad+\frac{1}{(t-t_0+i(x-x_0))^4}-\frac{1}{(t-t_0-i(x-x_0))^4}\Big]+\frac{1}{8\pi G}\partial_{t}\partial_{x}\delta(\boldsymbol{x}-\boldsymbol{x_0})\notag\\
    &\quad+\frac{i\mathcal{T}}{8\pi^2G}\Big[\frac{1}{(t-t_0+i(x+x_0))^2}-\frac{1}{(t-t_0-i(x+x_0))^2}\notag\\
    &\quad+\frac{1}{(t-t_0+i(x-x_0))^2}-\frac{1}{(t-t_0-i(x-x_0))^2}\Big],\notag\\
   \langle{T_{tx}(\boldsymbol{x})T_{tx}(\boldsymbol{x}_0)}\rangle&=\frac{3}{16\pi^2 G}\Big[\frac{1}{(t-t_0+i(x+x_0))^4}+\frac{1}{(t-t_0-i(x+x_0))^4}\notag\\
   &\quad-\frac{1}{(t-t_0+i|x-x_0|)^4}-\frac{1}{(t-t_0-i|x-x_0|)^4}\Big]-\frac{1}{8\pi G}\partial_{t}^2\delta(\boldsymbol{x}-\boldsymbol{x_0})\notag\\
	&\quad+\frac{\mathcal{T}}{8\pi^2G}\Big[\frac{1}{(t-t_0+i(x+x_0))^2}+\frac{1}{(t-t_0-i(x+x_0))^2}\notag\\
    &\quad-\frac{1}{(t-t_0+i(x-x_0))^2}-\frac{1}{(t-t_0-i(x-x_0))^2}\Big]-\frac{\mathcal{T}}{4\pi G}\delta(\boldsymbol{x}-\boldsymbol{x_0}),\notag\\
 \langle{T_{tx}(\boldsymbol{x})T_{xx}(\boldsymbol{x}_0)}\rangle&=\frac{-3i}{16\pi^2 G}\Big[\frac{1}{(t-t_0+i(x+x_0))^4}-\frac{1}{(t-t_0-i(x+x_0))^4}\notag\\
	&\quad+\frac{1}{(t-t_0+i(x-x_0))^4}-\frac{1}{(t-t_0-i(x-x_0))^4}\Big]\notag\\
    &\quad-\frac{i\mathcal{T}}{8\pi^2G}\Big[\frac{1}{(t-t_0+i(x+x_0))^2}-\frac{1}{(t-t_0-i(x+x_0))^2}\notag\\
    &\quad+\frac{1}{(t-t_0+i(x-x_0))^2}-\frac{1}{(t-t_0-i(x-x_0))^2}\Big].
    \end{align}
	The other two-point correlators can be computed using the conservation equation (\ref{conservation equation 1}) and the Bose symmetry of the correlators,
\begin{align}
		\langle{T_{tt}(\boldsymbol{x})T_{tt}(\boldsymbol{x}_0)}\rangle&=\frac{3}{16\pi^2 G}\Big[\frac{1}{(t-t_0+i(x+x_0))^4}+\frac{1}{(t-t_0-i(x+x_0))^4}\notag\\
    &\quad+\frac{1}{(t-t_0+i|x-x_0|)^4}+\frac{1}{(t-t_0-i|x-x_0|)^4}\Big]+\frac{1}{8\pi G}(\partial_{t}^2-\partial_{x}^2)\delta(\boldsymbol{x}-\boldsymbol{x}_0)\notag\\
    &\quad+\frac{\mathcal{T}}{8\pi^2G}\Big[\frac{1}{(t-t_0+i(x+x_0))^2}+\frac{1}{(t-t_0-i(x+x_0))^2}\notag\\
    &\quad+\frac{1}{(t-t_0+i|x-x_0|)^2}+\frac{1}{(t-t_0-i|x-x_0|)^2}\Big]+\frac{\mathcal{T}}{4\pi G}(1-2\partial_{t})\delta(\boldsymbol{x}-\boldsymbol{x}_0),\notag\\
\langle{T_{tt}(\boldsymbol{x})T_{xx}(\boldsymbol{x}_0)}\rangle&=-\frac{3}{16\pi^2 G}\Big[\frac{1}{(t-t_0+i(x+x_0))^4}+\frac{1}{(t-t_0-i(x+x_0))^4}\notag\\
    &\quad+\frac{1}{(t-t_0+i|x-x_0|)^4}+\frac{1}{(t-t_0-i|x-x_0|)^4}\Big]-\frac{1}{8\pi G}\partial_{t}^2\delta(\boldsymbol{x}-\boldsymbol{x}_0)\notag\\
    &\quad-\frac{\mathcal{T}}{8\pi^2G}\Big[\frac{1}{(t-t_0+i(x+x_0))^2}+\frac{1}{(t-t_0-i(x+x_0))^2}\notag\\
    &\quad+\frac{1}{(t-t_0+i|x-x_0|)^2}+\frac{1}{(t-t_0-i|x-x_0|)^2}\Big]-\frac{\mathcal{T}}{4\pi G}(3-2\partial_{t})\delta(\boldsymbol{x}-\boldsymbol{x}_0),\notag\\
    \langle{T_{xx}(\boldsymbol{x})T_{xx}(\boldsymbol{x}_0)}\rangle&=\frac{3}{16\pi^2 G}\Big[\frac{1}{(t-t_0+i(x+x_0))^4}+\frac{1}{(t-t_0-i(x+x_0))^4}\notag\\
    &\quad+\frac{1}{(t-t_0+i|x-x_0|)^4}+\frac{1}{(t-t_0-i|x-x_0|)^4}\Big]\notag\\
    &\quad+\frac{\mathcal{T}}{8\pi^2G}\Big[\frac{1}{(t-t_0+i(x+x_0))^2}+\frac{1}{(t-t_0-i(x+x_0))^2}\notag\\
    &\quad+\frac{1}{(t-t_0+i|x-x_0|)^2}+\frac{1}{(t-t_0-i|x-x_0|)^2}\Big]+\frac{\mathcal{T}}{4\pi G}(5-2\partial_{t})\delta(\boldsymbol{x}-\boldsymbol{x}_0).
\end{align}
\section{Holographic correlators of crosscap CFT$_2$ on $\mathbb{RP}^2$}\label{section 3}
In this section, we extend our method to the case of two-dimensional CFTs on non-orientable surfaces (XCFTs in \cite{Wei:2024zez}). Our calculation is based on the model constructed in \cite{Wei:2024zez}.
\subsection{Holographic dual of crosscap CFT$_2$}\label{subsection 3.1}
Firstly, we review the basic construction of the holographic dual of crosscap $\mathrm{CFT}_2$ (XCFT$_2$) \cite{Maloney:2016gsg, Wei:2024zez}.\par
A simple fact in topology is that any non-orientable manifold $\Sigma_g$ can be represented as the $\mathbb{Z}_2$ quotient of its orientable double cover $\hat{\Sigma}_g$, formally expressed as $\Sigma_g=\hat{\Sigma}_g/\mathbb{Z}_2$. Additionally, any three-dimensional manifold with boundary $\Sigma_g$ can be obtained by taking the $\mathbb{Z}_2$ quotient of the three-dimensional manifold with boundary $\hat{\Sigma}_g$. Therefore, the basic idea for the holography on non-orientable surfaces is that the bulk saddles with boundary $\Sigma_g$ can be identified by looking for $\mathbb{Z}_2$-invariant saddles with boundary $\hat{\Sigma}_g$. However, geometries obtained in this manner exhibit singularities when there are fixed points of the $\mathbb{Z}_2$ action in the manifold before the identification \cite{Maloney:2016gsg}, which prevents them from being global solutions to Einstein's equations.\par
A prescription to resolve these singularities is to introduce EOW branes in bulk to encompass them \cite{Wei:2024zez}, resulting in a construction similar to $\mathrm{AdS}_3/\mathrm{BCFT}_2$. For an $\mathrm{XCFT}_2$ lives on a non-orientable closed surface $\Sigma$, which is the boundary of the three-dimensional bulk $\mathcal{M}$, the dual bulk action still consists of three parts as the same as (\ref{AdS3/BCFT2 bulk action}):
\begin{equation}
	I_\mathrm{bulk}=-\frac{1}{16\pi G}\int_\mathcal{M}\sqrt{\mathcal{G}}\left(\mathcal{R}+2\right)-\frac{1}{8\pi G}\int_Q\sqrt{h}(K-T)-\frac{1}{8\pi G}\int_\Sigma\sqrt{\gamma}(B-1),\label{RP2 gravitational action}
\end{equation}
where $h_{ab}$ and $\gamma_{ij}$ denote the induced metrics on the EOW brane $Q$ and conformal boundary $\Sigma$ respectively. The scalar extrinsic curvatures $K:=h^{ab}K_{ab}$ (for $Q$) and $B:=\gamma^{ij}B_{ij}$ (for $\Sigma$) are constructed from their respective extrinsic curvature tensors $K_{ab}$ and $B_{ij}$.\par
The real projective plane $\mathbb{RP}^2$ is the simplest example of non-orientable surfaces, which is the quotient of the sphere $S^2$ under the antipodal map. We first consider a holographic CFT living on the unit sphere to obtain its dual bulk saddle. In that case, the dominant saddle is (Euclidean) global AdS$_3$, which is described by the metric
\begin{equation}
	\dif s^2=\dif\eta^2+\sinh^2\eta\left(\dif\theta^2+\sin^2\theta\dif\phi^2\right),\label{RP2 bulk metric}
\end{equation}
where $\eta\in(0,\infty)$, $\theta\in\left[0,\pi\right]$, $\phi\in[0,2\pi)$ with $\phi\sim\phi+2\pi$. The AdS radius is set to 1 for simplicity. The bulk saddle with an $\mathbb{RP}^2$ boundary can be obtained by imposing the antipodal identification $(\theta,\phi)\sim(\pi-\theta,\phi+\pi)$ on the boundary $S^2$ and extending this identification into bulk. A fixed point in the bulk exists, whose location depends on the extension.
Here, we assume that the dominant saddle is constructed by the identification with a fixed point at $\eta=0$, imposing the exact antipodal identification on every constant $\eta$ sphere. \par
To resolve the singularity, an EOW brane $Q$ with a tension $T$ is introduced at $\eta=\eta_*$.  The bulk $\mathcal{M}$ is the region between the brane $Q$ and the conformal boundary $\Sigma$. Similar to the case of AdS$_3$/BCFT$_2$, we impose the Dirichlet boundary condition $\delta\gamma_{ij}|_{\Sigma}=0$ on the conformal boundary $\Sigma$ and Neumann boundary condition
\begin{equation}
	K_{ab}-Kh_{ab}=-Th_{ab}\label{Neumann boundary condition}
\end{equation}
on the EOW brane $Q$, while the latter determines the relation between the brane profile and tension:
\begin{equation}
	\eta_*=\mathrm{arccoth}(-T),\quad T<-1.
\end{equation}
The partition function and holographic correlators of the XCFT$_2$ on $\mathbb{RP}^2$ can then be investigated based on this model.
\begin{figure}[H]
    \centering
    \includegraphics[scale=0.25]{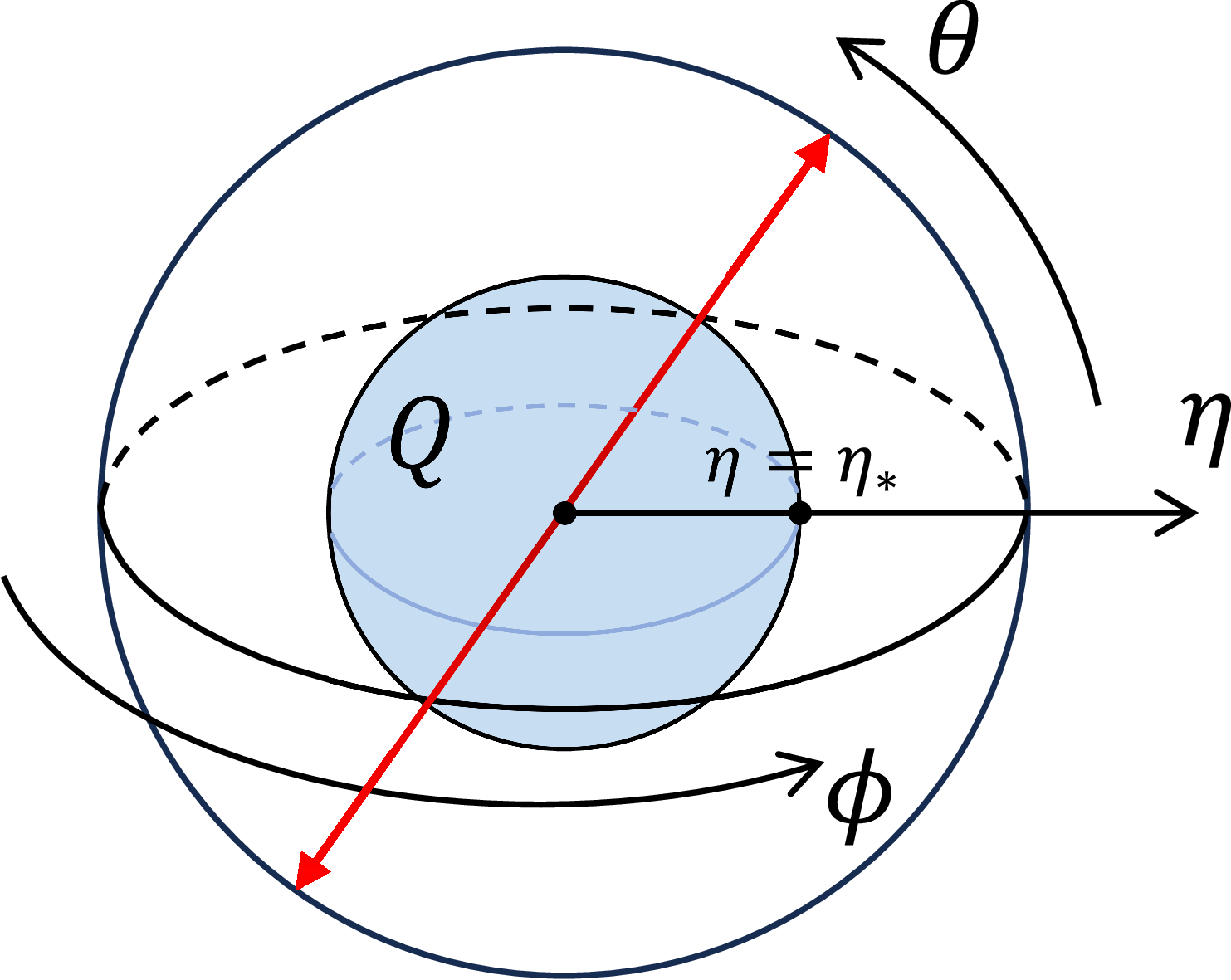}
    \caption{An illustration of the holographic dual of XCFT$_2$ on $\mathbb{RP}^2$. The EOW brane $Q$ is located at $\eta=\eta_*$. The red double arrow stands for the antipodal identification.}
    \label{holographic dual of RP2}
\end{figure}
\subsection{Holographic scalar correlators}
As a simple example, in this subsection, we use the standard GKPW relation to compute the exact one-point and two-point scalar correlators on $\mathbb{RP}^2$. We consider the following scalar field action in three-dimensional bulk,
\begin{align}
	 I_{\text{scalar}}=\frac{1}{2}\int_{\mathcal{M}}\text{d}^3x\sqrt{\mathcal{G}}\Big((\partial\Phi)^2+m^2\Phi^2\Big). \label{scalar bulk action}
\end{align}
In the semi-classical limit, scalar field action can be viewed as a perturbation of the gravitational action (\ref{RP2 gravitational action}). From now on, we will neglect the backreaction of the scalar field on the metric and calculate scalar correlators in the fixed background (\ref{RP2 bulk metric}). The scalar field equation of motion is 
\begin{align}
\frac{1}{\sqrt{\mathcal{G}}}\partial_{\mu}\Big(\sqrt{\mathcal{G}}\mathcal{G}^{\mu\nu}\partial_{\nu}\Phi\Big)-m^2\Phi=&0.\label{scalar EOM}
\end{align}
As discussed in \cite{Witten:1998qj,deHaro:2000vlm, Hung:2011ta}, this equation can be solved near the conformal boundary, and there are two homogeneous solutions for the scalar field $\Phi(\eta,\boldsymbol{x})$,
\begin{align}
\Phi(\eta,\boldsymbol{x})=&(2e^{-\eta})^{2-\Delta}\phi_{-}(\eta,\boldsymbol{x})+(2e^{-\eta})^{\Delta}\phi_+(\eta,\boldsymbol{x}),\label{expansion 1}
\end{align}
where $\Delta=1+\sqrt{1+m^2}$ is the scaling dimension of the dual operator, which has a lower bound $\Delta\geq 1$ called the BF bound \cite{Breitenlohner:1982bm, Breitenlohner:1982jf}. In (\ref{expansion 1}), both $\phi_{-}(\eta,\boldsymbol{x})$ and $\phi_+(\eta,\boldsymbol{x})$ can be expressed as Taylor series expansions in $e^{-2\eta}$. The leading order coefficient of each solution has a specific interpretation: the coefficient $\phi_{-}(\infty,\boldsymbol{x})$ (which is denoted by $\phi_{(0)}$ below) corresponds to the source coupled to the dual operator $O$, while the coefficient $\phi_{+}(\infty,\boldsymbol{x})$ (which is denoted by $\phi_{(2\Delta-2)}$ below) is related to one-point correlator in boundary CFT. Plugging (\ref{scalar EOM}) and (\ref{expansion 1}) into (\ref{scalar bulk action}), we find that the on-shall action diverges when evaluated at the conformal boundary. Following the prescription in \cite{deHaro:2000vlm}, we include the following counterterm to eliminate the divergence,
\begin{align}
    I_{\text{ct}}=\frac{2-\Delta}{2}\int_{\Sigma}\text{d}^2x\sqrt{\gamma}\Phi^2.
\end{align}
The renormalized one-point scalar correlator takes the form
\begin{align}
    \langle{O}\rangle&=-(2e^{-\eta})^{-\Delta}[\partial_{\eta}+(2-\Delta)]\Phi\Big|_{\eta\to\infty}\notag\\
    &=(2\Delta-2)\phi_{(2\Delta-2)}. \label{definition of O}
\end{align}\par
In the background (\ref{RP2 bulk metric}), the EOW brane $Q$ is located at $\eta=\eta_*$. We assume that the bulk scalar field $\Phi$ is coupled to the EOW brane by the following quadratic function \cite{Kastikainen:2021ybu},
\begin{align}
	I_{\text{brane}}=-\int_{Q}\text{d}^2x\sqrt{h}\Big(\lambda_1\Phi+\frac{1}{2}\lambda_2\Phi^2\Big),\label{quadratic coupling}
\end{align}
The variation of (\ref{scalar bulk action}) plus (\ref{quadratic coupling}) gives a Robin boundary condition for $\Phi$ \cite{mcavity1993quantum},
\begin{align}
	[\partial_{\eta}\Phi+\lambda_1+\lambda_2\Phi]\Big|_{Q}=0.\label{Robin boundary condition}
\end{align}\par
Let us first compute the one-point correlator $\langle{O}\rangle$, which is equivalent to finding the bulk solution of $\Phi$ with the Dirichlet boundary condition $\delta\Phi=0$ on $\Sigma$ and the Robin boundary condition (\ref{Robin boundary condition}) on $Q$. The scalar equation of motion (\ref{scalar EOM}) expressed in the coordinates $(\eta,\theta,\phi)$ takes the form
\begin{align}
    \Big[\partial_{\eta}^2+2\coth{\eta}\partial_{\eta}+\text{csch}^2\eta(\frac{1}{\sin{\theta}}\partial_\theta(\sin{\theta}\partial_\theta)+\frac{1}{\sin^2\theta}\partial_{\phi}^2)+\Delta(2-\Delta)\Big]\Phi=0.\label{RP2 EOM 1}
\end{align}
The solution can be decomposed by the spherical harmonics $\lbrace{Y_{lm}}\rbrace$,
\begin{align}
	\Phi(x,\theta,\phi)=\sum_{l\in 2\mathbb{N}^0}\sum_{m=-l}^{l}x^{\frac{l}{2}}(1-x)^{\frac{\Delta}{2}}y_{lm}(x)Y_{lm}(\theta,\phi),\label{decomposation}
\end{align}
where the new radial coordinate is defined as $x=\tanh^2\eta$. Since the solution on $\mathbb{RP}^2$ should be invariant under the involution $(\theta,\phi)\mapsto(\pi-\theta,\phi+\pi)$, the summation here only contains spherical harmonics with even $l$. Plugging (\ref{decomposation}) into (\ref{RP2 EOM 1}), we find that $y_{lm}$ satisfies the following hypergeometric equation,
\begin{align}
    x(1-x)y_{lm}''+[l+\frac{3}{2}-(l+\Delta+\frac{3}{2})x]y_{lm}'-\frac{(l+\Delta)(l+\Delta+1)}{4}y_{lm}=0.
\end{align}
From the Dirichlet boundary condition of $\Phi$ on $\Sigma$, we have
\begin{align}
	(1-x)^{\Delta-1}y_{lm}(x)\Big|_{x\to 1}=\int\text{d}\theta'\text{d}\phi'\sin{\theta'}Y^*_{lm}(\theta',\phi')\phi_{(0)}(\theta',\phi'). \label{DBC for y}
\end{align}
We assume the source $\phi_{(0)}$ is turned off in XCFT$_2$. Then, the right-hand side of (\ref{DBC for y}) vanishes and the general solution for $y_{lm}$ is
\begin{align}
    y_{lm}(x)=A_{lm}F(\frac{l+\Delta}{2},\frac{l+\Delta+1}{2},\Delta,1-x),\label{General solution for y}
\end{align}
where $F$ is the hypergeometric function \cite{abramowitz1968handbook}. From the Robin boundary condition (\ref{Robin boundary condition}) on $Q$, we have
\begin{align}
	2x^{\frac{1+l}{2}}(1-x)^{\frac{\Delta+2}{2}}(\frac{\text{d}}{\text{d}x}+\frac{l}{2x}+\frac{\Delta}{2(1-x)}+\frac{\lambda_2}{2\sqrt{x}(1-x)})y_{lm}\Big|_{x=\frac{1}{T^2}}+\lambda_1\sqrt{4\pi}\delta_{l,0}\delta_{m,0}=0.\label{RBC for y}
\end{align}
Plugging (\ref{General solution for y}) into (\ref{RBC for y}), we obtain
\begin{align}
	&A_{00}=\frac{-\lambda_1\sqrt{4\pi}(-T)^{\Delta+3}(T^2-1)^{-\frac{\Delta+2}{2}}}{\frac{\Delta+1}{2}F(\frac{\Delta+2}{2},\frac{\Delta+3}{2},1+\Delta,\frac{T^2-1}{T^2})-\frac{T^2(\Delta-\lambda_2T)}{T^2-1}F(\frac{\Delta}{2},\frac{\Delta+1}{2},\Delta,\frac{T^2-1}{T^2})},\notag\\
    &\quad\quad\quad\quad\quad\quad\quad\quad\quad  A_{lm}=0,\ \ \ \ \text{for }(l,m)\neq(0,0).\label{A00 coefficient}
\end{align}
Putting everything together and using the definition (\ref{definition of O}), we have
\begin{align}
	\langle{O}\rangle=\frac{(\Delta-1)A_{00}}{\sqrt{\pi}}.
\end{align}
Let us further consider the one-point correlator of heavy operators. Assume that the scaling dimension satisfies $c\gg \Delta\gg 1$. The hypergeometric functions in (\ref{A00 coefficient}) exhibit the following asymptotic behaviors,
\begin{align}
    F(\frac{\Delta}{2},\frac{\Delta+1}{2},\Delta,\frac{T^2-1}{T^2})&\sim \Big(\frac{2T}{T-1}\Big)^{\Delta},\notag\\
    F(\frac{\Delta+2}{2},\frac{\Delta+3}{2},\Delta+1,\frac{T^2-1}{T^2})&\sim \Big(\frac{2T}{T-1}\Big)^{\Delta}.
\end{align}
It follows that
\begin{align}
	\langle{O}\rangle&\sim\lambda_1e^{\Delta\log(\frac{1}{2}\sqrt{\frac{T-1}{T+1}})}=\frac{\lambda_1e^{\Delta\eta_*}}{2^{\Delta}}.
\end{align}
This result is consistent with the one-point correlator obtained by geodesic approximation in \cite{Wei:2024zez}.\par
To compute the scalar two-point correlator $\langle{OO}\rangle$, we perturb the boundary source $\phi_{(0)}$,
\begin{align}
	\phi_{(0)}(\theta,\phi)&=\epsilon\chi(\theta,\phi),
\end{align}
where $\epsilon$ is an infinitesimal parameter, and $\chi(\theta,\phi)$ is invariant under the involution $(\theta,\phi)\mapsto(\pi-\theta,\phi+\pi)$. The perturbed bulk scalar field can be written as a power series in $\epsilon$, $\Phi=\sum_{n=0}^{\infty}\epsilon^{n}\Phi^{[n]}$. Solving the scalar equation of motion (\ref{scalar EOM}) with boundary conditions (\ref{Robin boundary condition}) and (\ref{DBC for y}), and using the definition (\ref{definition of O}), we obtain the first-order variation
\begin{align}
    \langle{O(\theta,\phi)}\rangle^{[1]}&=(2\Delta-2)\sum_{l\in 2\mathbb{N}^0}\sum_{m=-l}^{l}A_{l}^{[1]}\int\text{d}\theta'\text{d}\phi'\sin{\theta'}Y^*_{lm}(\theta',\phi')\chi(\theta',\phi')Y_{lm}(\theta,\phi),
\end{align}
where
\begin{align}
	A_{l}^{[1]}&=\frac{(3\Delta+l(T^2-1)-2-\lambda_2T)G(l,\Delta;T)+\frac{(2+l-\Delta)(3+l-\Delta)}{2(\Delta-2)}G(l+1,\Delta-1;T)}{(\Delta T^2-l(T^2-1)-\lambda_2T^3)F(l,\Delta;T)-\frac{(T^2-1)(l+\Delta)(1+l+\Delta)}{2\Delta T^2}F(l+1,\Delta+1;T)}.
\end{align}
Here, we have used the notation
\begin{align}
	F(l,\Delta;T)&=F(\frac{l+\Delta}{2},\frac{l+\Delta+1}{2},\Delta,\frac{T^2-1}{T^2}),\notag\\
	G(l,\Delta;T)&=\Big(\frac{T^2-1}{T^2}\Big)^{1-\Delta}F(\frac{l-\Delta+2}{2},\frac{l-\Delta+3}{2},2-\Delta,\frac{T^2-1}{T^2}).
\end{align}
Finally, we obtain the two-point correlator $\langle{OO}\rangle$ from the standard GKPW relation,
\begin{align}
    \langle{O(\theta,\phi)O(\theta_0,\phi_0)}\rangle=&\frac{1}{\sqrt{g(\theta_0,\phi_0)}}\frac{\delta\langle{O(\theta,\phi)}\rangle^{[1]}}{\delta \chi(\theta_0,\phi_0)}\notag\\
    =&\frac{\Delta-1}{2\pi}\sum_{l\in 2\mathbb{N}^0}(2l+1)A^{[1]}_{l}P_{l}(\cos{\gamma}),
\end{align}
where $\cos{\gamma}=\cos{\theta}\cos{\theta_0}+\sin{\theta}\sin{\theta_0}\cos{(\phi-\phi_0)}$.

\subsection{Holographic stress tensor correlators}\label{3.3}
In this subsection, based on the method employed previously, we proceed to compute holographic stress tensor correlators on $\mathbb{RP}^2$.\par
After performing the coordinate transformation
\begin{equation}
	z=2e^{-\eta},
\end{equation}
the metric (\ref{RP2 bulk metric}) is transformed into the standard Fefferman-Graham form:
\begin{equation}
	\dif s^2=\frac{\dif z^2}{z^2}+\frac{1}{z^2}\left(1-\frac{z^2}{2}+\frac{z^4}{16}\right)\left(\dif\theta^2+\sin^2\theta\dif\phi^2\right),
\end{equation}
which is invariant under the antipodal map. According to the metric, we read off
\begin{equation}
	g^{(0)}=\begin{pmatrix}
		1&0\\0&\sin^2\theta
	\end{pmatrix},\quad g^{(2)}=\begin{pmatrix}
		-\frac{1}{2}&0\\0&-\frac{1}{2}\sin^2\theta
	\end{pmatrix},\quad g^{(4)}=\begin{pmatrix}
		\frac{1}{16}&0\\0&\frac{1}{16}\sin^2\theta
	\end{pmatrix}.
\end{equation}
It's easy to verify that they indeed satisfy Einstein's equations (\ref{g4 expression})(\ref{conservation equation 0})(\ref{trace relation 0}) in the Fefferman-Graham coordinates. Through (\ref{stress tensor one-point correlators}), we obtain the one-point correlators directly:
\begin{equation}
	\braket{T_{\theta\theta}}=\frac{1}{16\pi G},\quad\braket{T_{\phi\phi}}=\frac{1}{16\pi G}\sin^2\theta,\quad\braket{T_{\theta\phi}}=\braket{T_{\phi\theta}}=0,
\end{equation}
which satisfies the conservation law (\ref{conservation equation 1}) and the trace relation (\ref{trace relation 1}).\par
Our main aim is to compute two-point correlators of the stress tensor. Following the method utilized in section \ref{section 2}, we initiate by perturbing the boundary metric:
\begin{equation}
	\delta g_{(0)ij}\dif x^i\dif x^j=\epsilon\chi_{ij}\dif x^i\dif x^j,
\end{equation}
the variation of other Fefferman-Graham coefficients and one-point correlators can be formally written as
\begin{align}
	\delta g_{(2)ij}=\sum_{n=1}^\infty\epsilon^n g_{(2)ij}^{[n]},\quad\delta g_{(4)ij}=\sum_{n=1}^\infty\epsilon^n g_{(4)ij}^{[n]},\quad\delta\braket{T_{ij}}=\sum_{n=1}^\infty\epsilon^n\braket{T_{ij}}^{[n]}.
\end{align}
The perturbed brane profile is denoted as
\begin{equation}
    z(\theta,\phi)=z_*+\sum_{n=1}^\infty\epsilon^nf^{[n]}(\theta,\phi),
\end{equation}
where $z_*=2e^{-\eta_*}$ stands for the unperturbed brane profile. Similar to the case of AdS$_3$/BCFT$_2$ in the hyperbolic slicing coordinates, the Neumann boundary condition (\ref{Neumann boundary condition}) on the perturbed EOW brane provides a relationship between $g^{[1]}_{(2)ij}$ and $f^{[1]}$. In principle, by solving Einstein's equation with the boundary conditions order by order, we can derive any functional derivative which is of the form
\begin{equation}
	\frac{\delta\braket{T_{ij}}^{[n]}(\boldsymbol{x})}{\delta\chi_{i_1j_1}(\boldsymbol{x}_1)\delta\chi_{i_2j_2}(\boldsymbol{x}_2)\cdots\delta\chi_{i_nj_n}(\boldsymbol{x}_n)}.
\end{equation}
Thus, it's enough to obtain any holographic stress tensor correlators. However, the equations can be complicated, so we focus on calculating two-point correlators here.\par
Combining (\ref{stress tensor one-point correlators})(\ref{conservation equation 1})(\ref{trace relation 1}) and the Neumann boundary condition (\ref{Neumann boundary condition}), we obtain
\begin{align}
	&\braket{T_{\theta\theta}}^{[1]}=\frac{1}{16\pi Gz_*}\left(z_*\chi_{\theta\theta}-2f^{[1]}-2\csc^2\theta\partial_\phi^2f^{[1]}-2\cot\theta\partial_\theta f^{[1]}\right),\label{Tthetatheta1}\\
	&\braket{T_{\theta\phi}}^{[1]}=\frac{1}{16\pi Gz_*}\left(z_*\chi_{\theta\phi}-2\cot\theta\partial_\phi f^{[1]}+2\partial_\theta\partial_\phi f^{[1]}\right),\label{Tthetaphi1}\\
	&\braket{T_{\phi\phi}}^{[1]}=\frac{1}{16\pi Gz_*}\left(z_*\chi_{\phi\phi}-2\sin^2\theta f^{[1]}-2\sin^2\theta\partial_\theta^2f^{[1]}\right)\label{Tphiphi1}
\end{align}
and the equation that $f^{[1]}$ satisfies:
\begin{align}
	&\frac{1}{\sin\theta}\frac{\partial}{\partial\theta}\left(\sin\theta\frac{\partial f^{[1]}}{\partial\theta}\right)+\frac{1}{\sin^2\theta}\frac{\partial^2f^{[1]}}{\partial\phi^2}+2f^{[1]}-z_*\chi_{\theta\theta}-z_*\frac{1}{\sin^4\theta}\chi_{\phi\phi}+\frac{z_*}{2}\cot\theta\partial_\theta\chi_{\theta\theta}\nonumber\\
	&\quad\quad+z_*\frac{\cot\theta}{\sin^2\theta}\partial_\theta\chi_{\phi\phi}-\frac{z_*}{2\sin^2\theta}(\partial_\phi^2\chi_{\theta\theta}-2\partial_\theta\partial_\phi\chi_{\theta\phi}+\partial_\theta^2\chi_{\phi\phi})=0,\label{equation of brane variation}
\end{align}
For simplicity, we define
\begin{equation}
	f^{[1]}_1:=\dfrac{\delta f^{[1]}(\theta,\phi)}{\delta\chi_{\theta\theta}(\theta',\phi')},\quad f^{[1]}_2:=\dfrac{\delta f^{[1]}(\theta,\phi)}{\delta\chi_{\theta\phi}(\theta',\phi')},\quad f^{[1]}_3:=\dfrac{\delta f^{[1]}(\theta,\phi)}{\delta\chi_{\phi\phi}(\theta',\phi')}.
\end{equation}
Thus, following (\ref{equation of brane variation}), we have
\begin{align}
	&\frac{1}{\sin\theta}\frac{\partial}{\partial\theta}\left(\sin\theta\frac{\partial f^{[1]}_1}{\partial\theta}\right)+\frac{1}{\sin^2\theta}\frac{\partial^2f^{[1]}_1}{\partial\phi^2}+2f^{[1]}_1\nonumber\\
	&\hspace{3.6cm}=z_*\left(1-\frac{1}{2}\cot\theta\partial_\theta+\frac{1}{2\sin^2\theta}\partial_\phi^2\right)\delta_{\mathbb{RP}^2}^+(\theta,\phi;\theta',\phi'),\label{spherical harmonic 1}\\
	&\frac{1}{\sin\theta}\frac{\partial}{\partial\theta}\left(\sin\theta\frac{\partial f^{[1]}_2}{\partial\theta}\right)+\frac{1}{\sin^2\theta}\frac{\partial^2f^{[1]}_2}{\partial\phi^2}+2f^{[1]}_2=-\frac{z_*}{2\sin^2\theta}\partial_\theta\partial_\phi\delta_{\mathbb{RP}^2}^-(\theta,\phi;\theta',\phi'),\label{spherical harmonic 2}\\
	&\frac{1}{\sin\theta}\frac{\partial}{\partial\theta}\left(\sin\theta\frac{\partial f^{[1]}_3}{\partial\theta}\right)+\frac{1}{\sin^2\theta}\frac{\partial^2f^{[1]}_3}{\partial\phi^2}+2f^{[1]}_3\nonumber\\
	&\hspace{3.6cm}=z_*\left[\frac{1}{\sin^4\theta}+\frac{1}{2\sin^2\theta}\left(\partial_\theta^2-2\cot\theta\partial_\theta\right)\right]\delta_{\mathbb{RP}^2}^+(\theta,\phi;\theta',\phi')\label{spherical harmonic 3}
\end{align}
on the covering space $S^2$, where
\begin{align}
	\delta_{\mathbb{RP}^2}^+(\theta,\phi;\theta',\phi'):=&\ \frac{1}{2}\big[\delta(\theta-\theta')\delta(\phi-\phi')+\delta(\pi-\theta-\theta')\delta(\phi+\pi-\phi')\nonumber\\
	&\quad\quad\quad\quad+\delta(\theta-\pi+\theta')\delta(\phi-\phi'-\pi)+\delta(\theta'-\theta)\delta(\phi-\phi')\big]\nonumber\\
	=&\ \delta(\theta-\theta')\delta(\phi-\phi')+\delta(\theta+\theta'-\pi)\delta(\phi-\phi'-\pi),\\
	\delta_{\mathbb{RP}^2}^-(\theta,\phi;\theta',\phi'):=&\ \delta(\theta-\theta')\delta(\phi-\phi')-\delta(\theta+\theta'-\pi)\delta(\phi-\phi'-\pi)
\end{align}
appears in the functional derivative
\begin{equation}
	\frac{\delta\chi_{\theta\theta}(\theta,\phi)}{\delta\chi_{\theta\theta}(\theta',\phi')}=\frac{\delta\chi_{\phi\phi}(\theta,\phi)}{\delta\chi_{\phi\phi}(\theta',\phi')}=\delta_{\mathbb{RP}^2}^+(\theta,\phi;\theta',\phi'),\quad\frac{\delta\chi_{\theta\phi}(\theta,\phi)}{\delta\chi_{\theta\phi}(\theta',\phi')}=\frac{1}{2}\delta_{\mathbb{RP}^2}^-(\theta,\phi;\theta',\phi').
\end{equation}\par
Equipped with the $\mathbb{Z}_2$ invariance and regularity condition at $\theta=0,\pi$, equations (\ref{spherical harmonic 1})(\ref{spherical harmonic 2})(\ref{spherical harmonic 3}) can be solved straightforwardly:
\begin{align}
	f^{[1]}_1&=\frac{z_*(1+\tan^2\frac{\theta'}{2})}{16\pi\tan\frac{\theta'}{2}}\Big[e^{2i\phi'}\tan^2\frac{\theta'}{2}(\frac{1}{(e^{i\phi}\tan\frac{\theta}{2}-e^{i\phi'}\tan\frac{\theta'}{2})^2}+\frac{2e^{-i\phi}\tan\frac{\theta}{2}}{1+\tan^2\frac{\theta}{2}}\frac{1}{e^{i\phi}\tan\frac{\theta}{2}-e^{i\phi'}\tan\frac{\theta'}{2}})\notag\\
	&\quad+e^{-2i\phi'}\tan^2\frac{\theta'}{2}(\frac{1}{(e^{-i\phi}\tan\frac{\theta}{2}-e^{-i\phi'}\tan\frac{\theta'}{2})^2}+\frac{2e^{i\phi}\tan\frac{\theta}{2}}{1+\tan^2\frac{\theta}{2}}\frac{1}{e^{-i\phi}\tan\frac{\theta}{2}-e^{-i\phi'}\tan\frac{\theta'}{2}})\notag\\
	&\quad+e^{2i\phi'}\cot^2\frac{\theta'}{2}(\frac{1}{(e^{i\phi}\tan\frac{\theta}{2}+e^{i\phi'}\cot\frac{\theta'}{2})^2}+\frac{2e^{-i\phi}\tan\frac{\theta}{2}}{1+\tan^2\frac{\theta}{2}}\frac{1}{e^{i\phi}\tan\frac{\theta}{2}+e^{i\phi'}\cot\frac{\theta'}{2}})\notag\\
	&\quad+e^{-2i\phi'}\cot^2\frac{\theta'}{2}(\frac{1}{(e^{-i\phi}\tan\frac{\theta}{2}+e^{-i\phi'}\cot\frac{\theta'}{2})^2}+\frac{2e^{i\phi}\tan\frac{\theta}{2}}{1+\tan^2\frac{\theta}{2}}\frac{1}{e^{-i\phi}\tan\frac{\theta}{2}+e^{-i\phi'}\cot\frac{\theta'}{2}})\Big]\notag\\
	&\quad+\frac{z_*}{4}\delta_{\mathbb{RP}^2}^+(\theta,\phi;\theta',\phi'),\label{f11}
\end{align}
\begin{align}
	f^{[1]}_2&=\frac{iz_*(1+\tan^2\frac{\theta'}{2})^2}{16\pi}\Big[e^{2i\phi'}(\frac{1}{(e^{i\phi}\tan\frac{\theta}{2}-e^{i\phi'}\tan\frac{\theta'}{2})^2}+\frac{2e^{-i\phi}\tan\frac{\theta}{2}}{1+\tan^2\frac{\theta}{2}}\frac{1}{e^{i\phi}\tan\frac{\theta}{2}-e^{i\phi'}\tan\frac{\theta'}{2}})\notag\\
	&\quad-e^{-2i\phi'}(\frac{1}{(e^{-i\phi}\tan\frac{\theta}{2}-e^{-i\phi'}\tan\frac{\theta'}{2})^2}+\frac{2e^{i\phi}\tan\frac{\theta}{2}}{1+\tan^2\frac{\theta}{2}}\frac{1}{e^{-i\phi}\tan\frac{\theta}{2}-e^{-i\phi'}\tan\frac{\theta'}{2}})\notag\\
	&\quad-e^{2i\phi'}\cot^4\frac{\theta'}{2}(\frac{1}{(e^{i\phi}\tan\frac{\theta}{2}+e^{i\phi'}\cot\frac{\theta'}{2})^2}+\frac{2e^{-i\phi}\tan\frac{\theta}{2}}{1+\tan^2\frac{\theta}{2}}\frac{1}{e^{i\phi}\tan\frac{\theta}{2}+e^{i\phi'}\cot\frac{\theta'}{2}})\notag\\
	&\quad+e^{-2i\phi'}\cot^4\frac{\theta'}{2}(\frac{1}{(e^{-i\phi}\tan\frac{\theta}{2}+e^{-i\phi'}\cot\frac{\theta'}{2})^2}+\frac{2e^{i\phi}\tan\frac{\theta}{2}}{1+\tan^2\frac{\theta}{2}}\frac{1}{e^{-i\phi}\tan\frac{\theta}{2}+e^{-i\phi'}\cot\frac{\theta'}{2}})\Big],
\end{align}
\begin{align}
	f^{[1]}_3&=-\frac{z_*(1+\tan^2\frac{\theta'}{2})^3}{64\pi\tan^3\frac{\theta'}{2}}\Big[e^{2i\phi'}\tan^2\frac{\theta'}{2}(\frac{1}{(e^{i\phi}\tan\frac{\theta}{2}-e^{i\phi'}\tan\frac{\theta'}{2})^2}+\frac{2e^{-i\phi}\tan\frac{\theta}{2}}{1+\tan^2\frac{\theta}{2}}\frac{1}{e^{i\phi}\tan\frac{\theta}{2}-e^{i\phi'}\tan\frac{\theta'}{2}})\notag\\
	&\quad+e^{-2i\phi'}\tan^2\frac{\theta'}{2}(\frac{1}{(e^{-i\phi}\tan\frac{\theta}{2}-e^{-i\phi'}\tan\frac{\theta'}{2})^2}+\frac{2e^{i\phi}\tan\frac{\theta}{2}}{1+\tan^2\frac{\theta}{2}}\frac{1}{e^{-i\phi}\tan\frac{\theta}{2}-e^{-i\phi'}\tan\frac{\theta'}{2}})\notag\\
	&\quad+e^{2i\phi'}\cot^2\frac{\theta'}{2}(\frac{1}{(e^{i\phi}\tan\frac{\theta}{2}+e^{i\phi'}\cot\frac{\theta'}{2})^2}+\frac{2e^{-i\phi}\tan\frac{\theta}{2}}{1+\tan^2\frac{\theta}{2}}\frac{1}{e^{i\phi}\tan\frac{\theta}{2}+e^{i\phi'}\cot\frac{\theta'}{2}})\notag\\
	&\quad+e^{-2i\phi'}\cot^2\frac{\theta'}{2}(\frac{1}{(e^{-i\phi}\tan\frac{\theta}{2}+e^{-i\phi'}\cot\frac{\theta'}{2})^2}+\frac{2e^{i\phi}\tan\frac{\theta}{2}}{1+\tan^2\frac{\theta}{2}}\frac{1}{e^{-i\phi}\tan\frac{\theta}{2}+e^{-i\phi'}\cot\frac{\theta'}{2}})\Big]\notag\\
	&\quad+\frac{z_*(1+\tan^2\frac{\theta'}{2})^2}{16\tan^2\frac{\theta'}{2}}\delta_{\mathbb{RP}^2}^+(\theta,\phi;\theta',\phi').
\end{align}
With these solutions, we can take the functional derivative of (\ref{Tthetatheta1})(\ref{Tthetaphi1})(\ref{Tphiphi1}) with respect to the metric perturbation without any obstacles.\par
For example, we have
\begin{align}
	&\braket{T_{\theta\theta}(\theta,\phi)T_{\theta\theta}(\theta',\phi')}=\frac{-2}{\sqrt{g_{(0)}(\theta',\phi')}}\frac{\delta\braket{T_{\theta\theta}(\theta,\phi)}}{\delta g_{(0)}^{\theta\theta}(\theta',\phi')}+\frac{g_{(0)\theta\theta}(\theta,\phi)}{\sqrt{g_{(0)}(\theta',\phi')}}\braket{T_{\theta\theta}(\theta,\phi)}\delta^+_{\mathbb{RP}^2}(\theta,\phi;\theta',\phi')\nonumber\\
	&=\frac{2}{\sqrt{g_{(0)}(\theta',\phi')}}\frac{\delta\braket{T_{\theta\theta}}^{[1]}(\theta,\phi)}{\delta\chi_{\theta\theta}(\theta',\phi')}+\frac{1}{16\pi G\sin\theta'}\delta^+_{\mathbb{RP}^2}(\theta,\phi;\theta',\phi')\nonumber\\
	&=\frac{3e^{2i(\phi-\phi')}}{256\pi^2 G}\Big[\frac{\csc^4\frac{\theta}{2}\csc^4\frac{\theta'}{2}}{\left(1+e^{i(\phi-\phi')}\cot\frac{\theta}{2}\cot\frac{\theta'}{2}\right)^4}+\frac{\csc^4\frac{\theta}{2}\sec^4\frac{\theta'}{2}}{\left(1-e^{i(\phi-\phi')}\cot\frac{\theta}{2}\tan\frac{\theta'}{2}\right)^4}\notag\\
	&\quad+\frac{\sec^4\frac{\theta}{2}\csc^4\frac{\theta'}{2}}{\left(1-e^{i(\phi-\phi')}\tan\frac{\theta}{2}\cot\frac{\theta'}{2}\right)^4}+\frac{\sec^4\frac{\theta}{2}\sec^4\frac{\theta'}{2}}{\left(1+e^{i(\phi-\phi')}\tan\frac{\theta}{2}\tan\frac{\theta'}{2}\right)^4}\Big]+\text{contact terms}.
\end{align}
This is a result written on the covering space $S^2$. One can easily verify that it is invariant under the involution, and exhibit Bose symmetry clearly.\par
It is convenient to use complex coordinates to simplify the results. By taking
\begin{equation}
	w=e^{i\phi}\tan\frac{\theta}{2},\quad\bar{w}=e^{-i\phi}\tan\frac{\theta}{2},
\end{equation}
we present all the two-point correlators as follows:
\begin{align}
	\langle{T_{ww}(\boldsymbol{w})T_{ww}(\boldsymbol{w}')}\rangle&=\frac{3}{16\pi^2G}\frac{1}{(w-w')^4}-\frac{1}{16\pi G w'^4}\Big(\partial_w\partial_{\bar w}+\frac{2w'}{1+w'\bar w'}\partial_w\notag\\
	&\quad-\frac{2\bar w'}{1+w'\bar w'}\partial_{\bar w}-\frac{4w'\bar w'}{1+w'\bar w'}\Big)\delta^{(2)}(w+\bar w'^{-1}),\\
	\langle{T_{ww}(\boldsymbol{w})T_{\bar w\bar w}(\boldsymbol{w}')}\rangle&=\frac{3}{16\pi^2G}\frac{1}{(1+w\bar w')^4}-\frac{1}{16\pi G}\Big(\partial_w\partial_{\bar w}-\frac{2w'}{1+w'\bar w'}\partial_w\notag\\
	&\quad+\frac{2\bar w'}{1+w'\bar w'}\partial_{\bar w}-\frac{4}{1+w'\bar w'}\Big)\delta^{(2)}(w-w'),\\
	\langle{T_{w\bar w}(\boldsymbol{w})T_{ww}(\boldsymbol{w}')}\rangle&=\frac{1}{16\pi G}\Big(\partial_w^2-\frac{2\bar w'}{1+w'\bar w'}\partial_w\Big)\delta^{(2)}(w-w')\notag\\
	&\quad+\frac{1}{16\pi Gw'^4}\Big(\partial_{\bar w}^2+\frac{2w'(3+2w'\bar w')}{1+w'\bar w'}\partial_{\bar w}+\frac{2w'^2(3+w'\bar w')}{1+w'\bar w'}\Big)\delta^{(2)}(w+\bar w'^{-1}),\\
	\langle{T_{w\bar w}(\boldsymbol{w})T_{w\bar w}(\boldsymbol{w}')}\rangle&=-\frac{1}{16\pi G}\Big(\partial_w\partial_{\bar w}-\frac{2}{(1+w'\bar w')^2}\Big)\delta^{(2)}(w-w').
\end{align}

\subsection{Stress tensor correlators at finite cutoff}
At the end of this section, we investigate the holographic stress tensor one-point and two-point correlators in a cutoff AdS$_3$. We employ the Fefferman-Graham coordinates (\ref{Fefferman Graham coordinates}) in the bulk. Following the prescription in \cite{McGough:2016lol}, the Dirichlet boundary condition is imposed at the hard radial cutoff $z=z_c$. A natural holographic dictionary for cutoff AdS$_3$ is given by the generalized GKPW relation \cite{Hartman:2018tkw},
\begin{align}
	    Z_{\text{G}}[g_{(c)ij}]=\Big\langle{\text{exp}\Big[-\frac{1}{2}\int\text{d}^2x\sqrt{g_{(c)}}g^{ij}_{(c)}T_{ij}}\Big)\Big]\Big\rangle_{\text{EFT}},\label{generalized GKPW rekation}
\end{align}
where $g_{(c)ij}=g_{ij}(z_c,\boldsymbol{x})$ is the boundary metric on the cutoff surface. For a pure gravitational system in 3D spacetime, the dual EFT is obtained by $T\bar T$ deformation of the original CFT, which is defined by the following flow equation for the field theory action \cite{Zamolodchikov:2004ce, Smirnov:2016lqw},
\begin{align}
	\frac{\text{d}S_{\mu}}{\text{d}\mu}=-\frac{1}{4}\int\text{d}^2x\ \text{det}[T_{\mu}],
\end{align}
where the deformation parameter\begin{align}
	\mu=16\pi G z_c^2.\label{deformation parameter}
\end{align}
\subsubsection{Deformed one-point correlators}
We begin with the Euclidean global AdS$_3$ with the metric (\ref{RP2 bulk metric}). Using the following coordinate transformation \cite{Tian:2023fgf},
\begin{align}
	z=2r_0e^{-\eta},\ \ \ w=e^{i\phi}\tan\frac{\theta}{2},\ \ \ \bar w=e^{-i\phi}\tan\frac{\theta}{2},
\end{align}
we have
\begin{align}
\text{d}s^2=\frac{\text{d}z^2}{z^2}+\frac{1}{z^2}\Big(1-\frac{z^2}{2r_0^2}+\frac{z^4}{16r_0^4}\Big)\frac{4r_0^2\text{d}w\text{d}\bar w}{(1+w\bar w)^2}. \label{bulk metric with complex sphere}
\end{align}
Now, let us move the Dirichlet boundary to $z=z_c=\sqrt{\frac{\mu}{16\pi G}}$. The boundary metric on this radial slice still takes the form of the metric on the sphere,
\begin{align}
	g_{(c)ij}\text{d}x^i\text{d}x^j=\frac{4r_{\mu}^2\text{d}w\text{d}\bar w}{(1+w\bar w)^2},
\end{align}
with
\begin{align}
	r_0^2=\frac{1}{4}\Big[\frac{\mu}{16\pi G}+2r_{\mu}^2\Big(1+\sqrt{1+\frac{\mu}{16\pi Gr_{\mu}^2}}\Big)\Big]\label{deformed radial}
\end{align}
Here $r_{\mu}$ is the radial of $\mathbb{RP}^2$ for which the deformed field theory lives. From now on we will set $r_{\mu}=1$. The Brown-York tensor on the cutoff boundary is defined as 
\begin{align}
	\langle{T_{ij}}\rangle_{\mu}=-\frac{1}{8\pi G}(K_{(c)ij}-K_{(c)}\gamma_{(c)ij}+\gamma_{(c)ij}),\label{cutoff Brown-York tensor}
\end{align}
where $\gamma_{(c)ij}$ and $K_{(c)ij}$ are the induced metric and the extrinsic curvature at $z=z_c$, respectively. From the bulk Einstein's equation, one can find that $\langle{T_{ij}}\rangle_{\mu}$ satisfies the conservation law and the deformed trace relation \cite{Li:2020zjb, Guica:2019nzm},
\begin{align}
	\nabla_{(c)}^{i}\langle{T_{ij}}\rangle_{\mu}&=0,\notag\\
	g^{ij}_{(c)}\langle{T_{ij}}\rangle_{\mu}&=\frac{1}{16\pi G}R_{(c)}-\frac{\mu}{2}\text{det}[\langle{T}\rangle_\mu].\label{deformed conservation law and trace relation}
\end{align}
 Plugging (\ref{bulk metric with complex sphere}) and (\ref{deformed radial}) into (\ref{cutoff Brown-York tensor}), we obtain
\begin{align}
	\langle{T_{w\bar w}}\rangle_{\mu}=-\frac{4}{\mu}\Big(1-\sqrt{1+\frac{\mu}{16\pi G}}\Big)\frac{1}{(1+w\bar w)^2},\ \ \ \langle{T_{ww}}\rangle_{\mu}=\langle{T_{\bar w\bar w}}\rangle_{\mu}=0. \label{deformed one-point correlators}
\end{align}
The results are consistent with the deformed one-point correlators on the sphere \cite{Li:2020pwa}, as the latter have already been $\mathbb{Z}_2$-invariant.
\subsubsection{Deformed two-point correlators}
The deformed two-point correlators can be computed using the approach described in subsection \ref{3.3}. Firstly, we perturb the boundary metric on the cutoff surface,
\begin{align}
	\delta g_{(c)ij}(w,\bar w)=\epsilon\chi_{ij}(w,\bar w).\label{cutoff boundary metric perturbation}
\end{align}
The perturbed Brown-York tensor on the cutoff surface can be written as a power series in $\epsilon$,
\begin{align}
	\langle{T_{ij}(\epsilon;\boldsymbol{x})}\rangle_{\mu}=\sum_{n=0}^{\infty}\epsilon^n\langle{T_{ij}(\boldsymbol{x})}\rangle_{\mu}^{[n]}.\label{cutoff one-point correlator perturbation}
\end{align}
Plugging (\ref{cutoff boundary metric perturbation}) and (\ref{cutoff one-point correlator perturbation}) into (\ref{deformed conservation law and trace relation}), and extracting the coefficients of order $\epsilon^1$, we obtain
\begin{align}
\partial_{\bar w}\langle{T_{ww}}\rangle^{[1]}_{\mu}&=-(\partial_{w}+\frac{2\bar w}{1+w\bar w})\langle{T_{w\bar w}}\rangle^{[1]}_{\mu}\notag\\
&\quad-\frac{2(1-\sqrt{1+\frac{\mu}{16\pi G}})}{\mu(1+w\bar w)}\Big[(1+w\bar w)(\partial_{\bar w}\chi_{ww}+\partial_w\chi_{w\bar w})+2\bar w\chi_{w\bar w}\Big],\label{2-pt deformed conservation equation 1}\\
\partial_{w}\langle{T_{\bar w\bar w}}\rangle^{[1]}_{\mu}&=-(\partial_{\bar w}+\frac{2w}{1+w\bar w})\langle{T_{w\bar w}}\rangle^{[1]}_{\mu}\notag\\
&\quad-\frac{2(1-\sqrt{1+\frac{\mu}{16\pi G}})}{\mu(1+w\bar w)}\Big[(1+w\bar w)(\partial_{\bar w}\chi_{w\bar w}+\partial_w\chi_{\bar w\bar w})+2w\chi_{w\bar w}\Big],\label{2-pt deformed conservation equation 2}\\
	\langle{T_{w\bar w}}\rangle^{[1]}_{\mu}&=-\frac{1}{16\pi G\sqrt{1+\frac{\mu}{16\pi G}}}\Big[\Big(1+2w\bar w-\frac{\mu}{16\pi G(1+\sqrt{1+\frac{\mu}{16\pi G}})^2}\Big)\chi_{w\bar w}\notag\\
	&\quad-\frac{1+w\bar w}{4}\Big((2w\partial_{\bar w}+(1+w\bar w)\partial_{\bar w}^2)\chi_{ww}-2(2w\partial_w+2\bar w\partial_{\bar w}\notag\\
	&\quad+(1+w\bar w)\partial_{w}\partial_{\bar w})\chi_{w\bar w}+(2\bar w\partial_{w}+(1+w\bar w)\partial_w^2)\chi_{\bar w\bar w}\Big)\Big].\label{2-pt deformed trace relation}
\end{align}
Meanwhile, the perturbed bulk metric is written in a specific Fefferman-Graham coordinate system (\ref{Fefferman Graham coordinates}). The Fefferman-Graham coefficients (at the first-order in $\epsilon$) in (\ref{Fefferman-Graham expansion}) can be expressed as
\begin{align}
	g^{[1]}_{(0)ww}&=\frac{1}{4}(2+\frac{3\mu}{16\pi G}+2\sqrt{1+\frac{\mu}{16\pi G}})\chi_{ww}-\frac{\mu}{4}(1+\sqrt{1+\frac{\mu}{16\pi G}})\langle{T_{ww}}\rangle^{[1]}_{\mu},\notag\\
	g^{[1]}_{(0)w\bar w}&=\frac{1}{4}(2-\frac{\mu}{16\pi G}+2\sqrt{1+\frac{\mu}{16\pi G}})\chi_{w\bar w}+\frac{\mu}{4}(1+\sqrt{1+\frac{\mu}{16\pi G}})\langle{T_{w\bar w}}\rangle^{[1]}_{\mu},\notag\\
	g^{[1]}_{(0)\bar w\bar w}&=\frac{1}{4}(2+\frac{3\mu}{16\pi G}+2\sqrt{1+\frac{\mu}{16\pi G}})\chi_{\bar w\bar w}-\frac{\mu}{4}(1+\sqrt{1+\frac{\mu}{16\pi G}})\langle{T_{\bar w\bar w}}\rangle^{[1]}_{\mu},\notag\\
	g^{[1]}_{(2)ww}&=-\frac{8\pi G}{\mu}(2+\frac{3\mu}{16\pi G}-2\sqrt{1+\frac{\mu}{16\pi G}})\chi_{ww}+8\pi G\sqrt{1+\frac{\mu}{16\pi G}}\langle{T_{ww}}\rangle^{[1]}_{\mu},\notag\\
		g^{[1]}_{(2)w\bar w}&=\frac{8\pi G}{\mu}(2+\frac{\mu}{16\pi G}-2\sqrt{1+\frac{\mu}{16\pi G}})\chi_{w\bar w}-8\pi G\sqrt{1+\frac{\mu}{16\pi G}}\langle{T_{w\bar w}}\rangle^{[1]}_{\mu},\notag\\
			g^{[1]}_{(2)\bar w\bar w}&=-\frac{8\pi G}{\mu}(2+\frac{3\mu}{16\pi G}-2\sqrt{1+\frac{\mu}{16\pi G}})\chi_{\bar w\bar w}+8\pi G\sqrt{1+\frac{\mu}{16\pi G}}\langle{T_{\bar w\bar w}}\rangle^{[1]}_{\mu}.\label{deformed FG coefficients}
\end{align}
The coefficient $g_{(4)ij}$ is determined by $g_{(0)ij}$ and $g_{(2)ij}$ according to (\ref{g4 expression}). Then, we can construct the perturbed bulk metric using the boundary metric variation $\chi_{ij}$ and the deformed Brown-York tensor $\langle{T_{ij}}\rangle^{[1]}_{\mu}$. On the other hand, the perturbed bulk metric should satisfy the Neumann boundary condition (\ref{Neumann boundary condition BCFT2}) on the EOW brane $Q$. The brane profile\footnote{Here we assume that the brane tension $T$ does not flow.} is
\begin{align}
    Q:\ \ \ z(w,\bar w)=z^*_{\mu}+\sum_{n=1}^{\infty}\epsilon^n f_{\mu}^{[n]}(w,\bar w),\label{deformed new profile}
\end{align}
where
\begin{align}
	z^*_{\mu}=\Big(1+\sqrt{1+\frac{\mu}{16\pi G}}\Big)\sqrt{\frac{T+1}{T-1}}.
\end{align}
By substituting the perturbed bulk metric and the deformed brane profile (\ref{deformed new profile}) into the Neumann boundary condition (\ref{Neumann boundary condition BCFT2}), we obtain the following expressions,
\begin{align}
	\langle{T_{ww}}\rangle^{[1]}_{\mu}&=\frac{1}{8\pi Gz^*_{\mu}}(\partial_w^2+\frac{2\bar w}{1+w\bar w}\partial_{w})f^{[1]}_{\mu}-\frac{2}{\mu}(1-\sqrt{1+\frac{\mu}{16\pi G}})\chi_{ww},\notag\\
	\langle{T_{w\bar w}}\rangle^{[1]}_{\mu}&=-\frac{1}{8\pi Gz^*_{\mu}}(\partial_{w}\partial_{\bar w}+\frac{2}{(1+w\bar w)^2})f^{[1]}_{\mu}-\frac{2}{\mu}(1-\sqrt{1+\frac{\mu}{16\pi G}})\chi_{w\bar w},\notag\\
	\langle{T_{\bar w\bar w}}\rangle^{[1]}_{\mu}&=\frac{1}{8\pi Gz^*_{\mu}}(\partial_{\bar w}^2+\frac{2w}{1+w\bar w}\partial_{\bar w})f^{[1]}_{\mu}-\frac{2}{\mu}(1-\sqrt{1+\frac{\mu}{16\pi G}})\chi_{\bar w\bar w}. \label{deformed BY tensor in terms of brane profile}
\end{align}
Plugging (\ref{deformed BY tensor in terms of brane profile}) into (\ref{2-pt deformed conservation equation 1}) and (\ref{2-pt deformed conservation equation 2}), we find that the conservation law is automatically satisfied. Moreover, from the trace relation (\ref{2-pt deformed trace relation}), we have
\begin{align}
	(\partial_{w}\partial_{\bar w}+\frac{2}{(1+w\bar w)^2})f^{[1]}_{\mu}&=-\frac{z^*_{\mu}(1+w\bar w)}{8\sqrt{1+\frac{\mu}{16\pi G}}}\Big[(2w\partial_{\bar w}+(1+w\bar w)\partial_{\bar w}^2)\chi_{ww}\notag\\
	&\quad-2(4+2w\partial_w+2\bar w\partial_{\bar w}+(1+w\bar w)\partial_{w}\partial_{\bar w})\chi_{w\bar w}\notag\\
	&\quad+(2\bar w\partial_{w}+(1+w\bar w)\partial_{w}^2)\chi_{\bar w\bar w}\Big].
\end{align}
The solution can be written as
\begin{align}
	f^{[1]}_{\mu}(\boldsymbol{w})=\frac{1+\sqrt{1+\frac{\mu}{16\pi G}}}{2\sqrt{1+\frac{\mu}{16\pi G}}}f^{[1]}_{0}(\boldsymbol{w}).\label{solution of deformed brane profile}
\end{align}
Finally, by plugging (\ref{solution of deformed brane profile}) into (\ref{deformed BY tensor in terms of brane profile}) and using the definition (\ref{holographic correlator}), we can express the deformed two-point correlators as
\begin{align}
	\langle{T_{ij}(\boldsymbol{w})T^{kl}(\boldsymbol{w}')}\rangle_{\mu}&=\frac{1}{\sqrt{1+\frac{\mu}{16\pi G}}}\Big[\langle{T_{ij}(\boldsymbol{w})T^{kl}(\boldsymbol{w}')}\rangle_{\text{XCFT}}\notag\\
	&\quad+\frac{\mu(1+w'\bar w')^2}{256\pi^2G^2(1+\sqrt{1+\frac{\mu}{16\pi G}})^2}\Big(\frac{\delta g_{ij}(\boldsymbol{w})}{\delta g_{kl}(\boldsymbol{w}')}+\frac{g_{ij}g^{\alpha\beta}}{2}\frac{\delta g_{\alpha\beta}(\boldsymbol{w})}{\delta g_{kl}(\boldsymbol{w}')}\Big)\Big],
\end{align}
where 
\begin{align}
    &\frac{\delta g_{ww}(\boldsymbol{w})}{\delta g_{ww}(\boldsymbol{w}')}=\delta^{(2)}(w-w'),\ \ \ \frac{\delta g_{ww}(\boldsymbol{w})}{\delta g_{\bar w\bar w}(\boldsymbol{w}')}=\frac{\bar w^2}{w^2}\delta^{(2)}(w+w'^{-1}),\notag\\
    &\frac{\delta g_{w\bar w}(\boldsymbol{w})}{\delta g_{w\bar w}(\boldsymbol{w}')}=\frac{\delta g_{w\bar w}(\boldsymbol{w})}{\delta g_{\bar ww}(\boldsymbol{w}')}=\frac{1}{2}\delta^{(2)}(w-w').
\end{align}
\section{Conclusion and outlook}
This paper investigates the holographic correlators of BCFT$_2$ and the crosscap CFT$_2$ on $\mathbb{RP}^2$. Our calculations are based on the standard GKPW relation and employ the semiclassical approximation. Firstly, we examine the stress tensor correlators within the framework of AdS$_3$/BCFT$_2$. For the case of a tensionless brane, we obtain exact two-point and three-point correlators and derive some recurrence relations for computing higher-point correlators. Moreover, we switch to the hyperbolic slicing coordinates to compute the two-point and three-point stress tensor correlators with a general brane tension. Our recurrence relations are consistent with the Ward identity in BCFT$_2$, thus providing a concrete verification of AdS$_3$/BCFT$_2$ correspondence. Secondly, we employ the holographic prescription in \cite{Wei:2024zez} to investigate the correlators of crosscap CFT$_2$ on $\mathbb{RP}^2$. We first obtain the exact one-point and two-point scalar correlators. In the limit of large conformal dimension, our one-point correlator aligns with that obtained by the geodesic approximation. Furthermore, we calculate the holographic two-point correlators of stress tensor at conformal infinity and a finite cutoff.\par
There are some remaining questions and interesting future directions. It is important to notice that the correlators obtained in section \ref{section 2} only include the stress tensor operators living in the bulk of BCFT$_2$. From a field theory perspective, another set of operators lives on the boundary of BCFT$_2$, which cannot be obtained by moving the bulk operators to the vicinity of the boundary. The holographic constructions of these boundary operators have been extensively investigated in \cite{Geng:2021iyq, Kawamoto:2022etl, Kusuki:2021gpt, Kusuki:2022wns, Kusuki:2022bic, Miyaji:2022dna, Biswas:2022xfw, Kusuki:2022ozk}. Extending our calculations to these holographic constructions and including the correlators of boundary operators would be an interesting future direction.\par
In subsection \ref{subsection 2.3}, we switch to the hyperbolic slicing coordinates and compute the stress tensor correlators with a non-zero brane tension. The corresponding calculations can be readily extended to other contexts, such as stress tensor correlators on $\mathbb{RP}^2$, discussed in subsection \ref{3.3}. One of the crucial points is that the global bulk metric can be expressed by the near boundary solution (which consists of the first two Fefferman-Graham coefficients $g_{(0)}$ and $g_{(2)}$). However, this point is not applicable in higher dimensions because the Fefferman-Graham expansion contains infinite terms in higher-dimensional spacetime. Developing a methodological approach to compute the holographic stress tensor correlators in higher dimensions is necessary.\par
The holographic stress tensor correlators in AdS/BCFT have several other promising directions for future exploration. Firstly, we can introduce a scalar field in the AdS bulk and consider its back-reaction to the geometry. A well-known solution in this setup is the Janus solution, which has been investigated in many works \cite{Bak:2003jk, Freedman:2003ax, Papadimitriou:2004rz, Bak:2007jm, Chiodaroli:2009yw, Bak:2011ga, Chiodaroli:2016jod, Auzzi:2021nrj, Suzuki:2022xwv}. Besides, we can introduce a brane-localized scalar field on the EOW brane \cite{Erdmenger:2014xya, Kanda:2023zse, Liu:2024oxg, Fujiki:2025yyf} and study its impact on holographic correlators. Recently, the holographic aspects of the cutoff AdS$_3$/BCFT$_2$ have been extensively investigated \cite{Deng:2023pjs, Deng:2024dct, Basu:2024xjq}. It would be intriguing to compute the stress tensor correlators in these contexts.\par
In section \ref{section 3}, we compute the holographic scalar correlators and stress tensor correlators of crosscap CFT$_2$ on $\mathbb{RP}^2$. An important future direction is to study the holographic correlators on other non-orientable surfaces, such as the Klein bottle $\mathbb{K}^2$. The Klein bottle can be represented as the $\mathbb{Z}_2$ quotient of a rectangular torus, $\mathbb{K}^2=T^2/\mathbb{Z}_2$. From the holographic perspective, two classical bulk saddles \cite{Maloney:2016gsg, Wei:2024zez} exist. The smooth one is the Euclidean geon geometry, which is obtained by taking the quotient of the non-rotating BTZ black hole \cite{Louko:1998hc}. The non-smooth one arises from the quotient of thermal AdS and contains two singularities in the bulk. For the non-smooth saddle, the author of \cite{Wei:2024zez} introduces two disconnected EOW branes to exclude the singularities, resulting in a finite bulk action that contributes to the holographic $\mathbb{K}^2$ partition function. Exploring the holographic aspects of $\mathbb{K}^2$ in this bulk construction is also worthwhile, which could provide valuable insights for the study of CFTs on the Klein bottle such as \cite{Tu:2017wks,Zhang:2022gtb}. A more general non-orientable surface can be represented as the connected sum of $N$ copies of the real projective plane, with its double cover being a genus-$(N-1)$ Riemann surface. As outlined in \cite{Maloney:2016gsg}, the bulk saddles for this non-orientable surface can be constructed by taking the $\mathbb{Z}_2$ quotient of the saddles for its double cover. It would be interesting to investigate the holographic partition function and correlation functions of a general non-orientable surface. Recently, a novel non-orientable AdS$_3$ spacetime has been constructed by the authors of \cite{Pathak:2024cpo}, which differs from the construction presented in subsection \ref{subsection 3.1}. An intriguing future direction involves computing the holographic correlators in this AdS$_3$ spacetime and investigating their correspondence in the dual field theory.

\section*{Acknowledgments}
We would like to thank Bin Chen, Jue Hou, Yi Li, Hao Ouyang, Yuan Sun and Hong-Hao Tu for their valuable discussions on this work. SH would like to appreciate the financial support from Ningbo University, the Max Planck Partner Group, and the Natural Science Foundation of China Grants (No.~12475053, No.~12075101, No.~12235016, No.~12347209).

\bibliographystyle{JHEP}
\bibliography{reference.bib}

\end{document}